\documentclass[11pt]{article}

\usepackage[margin=1in]{geometry}
\usepackage{subfig}
\usepackage{lastpage} 
\usepackage{fancyhdr}
\usepackage{multirow}

\usepackage{natbib}
\usepackage[english]{babel}
\usepackage[utf8]{inputenc}
\usepackage{amsfonts}
\usepackage{amsmath} 
\usepackage{amssymb}
\usepackage{graphicx}
\usepackage{amsthm}

\usepackage{enumerate}
\usepackage{float}
\usepackage{mdframed}
\usepackage{listings}
\usepackage{array}
\usepackage{bbm}
\usepackage[normalem]{ulem}
\usepackage[]{algorithm2e}
\usepackage{tablefootnote}

\usepackage[hidelinks]{hyperref}

\renewcommand{\hat}{\widehat}

\def\red#1{{\color{red} #1}}

\title{Identification of Parkinson's Disease Subtypes with Divisive Hierarchical Bayesian Clustering for Longitudinal and Time-to-Event Data}

\author{Elliot Burghardt, Daniel Sewell, Joseph Cavanaugh}
\date{}
\begin{document}

\maketitle

 \begin{abstract}
 In heterogeneous disorders like Parkinson’s disease (PD), differentiating the affected population into subgroups plays a key role in future research. Discovering subgroups can lead to improved treatments through more powerful enrichment of clinical trials, elucidating pathogenic mechanisms, and identifying biomarkers of progression and prognosis. Cluster analysis is a commonly used method to identify subgroups; however, cluster analysis methods are typically restricted to static data or temporal data of a single variable. Progression of a complex disease process may be more appropriately represented by several longitudinal and/or time-to-event variables. Clustering with longitudinal and time-to-event data presents challenges, such as correlations between clustering variables, temporal dependencies, missing data, and censoring. To address these challenges, we present Divisive Hierarchical Bayesian Clustering methods featuring models for multivariate longitudinal trajectories and semi-parametric models for survival data to identify subgroups in idiopathic PD with differing progression patterns from the Parkinson’s Progression Markers Initiative (PPMI) database.
\end{abstract}

\flushleft

\section{Introduction}
\label{sec:introduction}

PD is one of the most common neurodegenerative diseases in the world. \cite{tysnes2017epidemiology} estimate that PD affects 1\% of the global population above 60 years old. In 2018, \cite{marras2018prevalence} estimated that 930,000 individuals at least 45 years old in the United States would have PD by 2020 and over 1.2 million by 2030. PD often manifests as a combination of progressive motor symptoms (e.g. rest tremor, rigidity, bradykinesia, and postural instability) and non-motor symptoms (e.g. mood and anxiety disorders, sleep disturbances, cognitive, olfactory, and autonomic dysfunction). Progression of PD symptoms is highly variable, and there are currently no symptoms or signs that reliably predict an individual's future PD trajectory \citep{alves2006changes}. \newline

In addition to the burden that this uncertainty places on patients and families, the variability and unpredictability of disease trajectories and the lack of robust biomarkers impede the development of new and disease-modifying treatments \citep{marek2011parkinson}. There have been many potential disease-modifying agents that demonstrate success (a neuroprotective effect) in animal models, but nearly all have failed in clinical trials \citep{stocchi2014therapy, leonard2020genetic}. Thus, there are no treatments that have been established as disease-modifying in PD \citep{stocchi2014therapy, leonard2020genetic}. It is believed that the variability of the PD disease course and its pathogenesis may be attributed to the existence of disease subtypes \citep{thenganatt2014parkinson}. The identification of subgroups and their related markers\textemdash of diagnosis and prognosis\textemdash may be helpful in elucidating neurodegeneration mechanisms, developing appropriate therapies, and improving clinical trials that enrich for the target population \citep{thenganatt2014parkinson, leonard2020genetic, krasniqidata}. \newline

While previous studies have defined PD subgroups (e.g. tremor-dominant vs. postural instability and gait difficulty subtypes, or based upon a cutoff for the Unified Parkinson’s Disease Rating Scale), these methods concentrate on a single feature of PD, neglecting its multifaceted nature \citep{zhang2019data}. Because PD is associated with a broad range of symptoms, signs, genetic factors, and neuroimaging results as well as high variability in individual disease course, identifying comprehensive subgroups is challenging \citep{krasniqidata}. To address this challenge, machine learning methods may enhance our ability to detect subgroups when differences are not otherwise easily perceived. Cluster analysis is one tool for discovering subgroups in a population, uncovering latent patterns in data. 
However, traditional clustering methods are unlikely to capture the complexity of PD, which has a variable course across a variety of clinical characteristics and underlying mechanisms \citep{de2019deep}. Traditional clustering methods do not account for temporal dependencies between observations or correlations between variables. Using static baseline features to form clusters is a common approach because it removes the need to address temporal trends, and it could be used to establish subgroups at enrollment. However, identifying clusters through baseline measurements is typically only useful if these clusters experience diverging progression patterns. That is, differences between clusters over time is often of paramount interest, and this can only be evaluated post hoc when clusters are formed using baseline data.\newline

A variety of clustering approaches have been utilized to identify subgroups in PD. These methods typically fall into the following categories: those that use baseline data to form clusters \citep{emon2020clustering, wang2020association}, those that simplify the characterization of longitudinal data into slopes or distance matrices \citep{vavougios2018identification, krishnagopal2020identifying, zhang2019data, krasniqidata}, and those that rely on latent variables \citep{de2019deep}. Studies involving time-to-event data to identify clusters defining PD subgroups were not uncovered in a literature search. \cite{emon2020clustering} performed clustering using sparse autoencoders and sparse non-negative matrix factorization of baseline biological variables to identify mixed AD/PD subgroups. They tested for disease relevance between clusters identified via brain imaging (DaTSCAN and MRI) and clinical features and trascriptome and methylome profiles. \cite{wang2020association} used voxel-level neuroanatomic features that were significantly correlated with clinical scores (e.g. MDS-UPDRS) as inputs to traditional clustering methods, agglomerative clustering and k-means. \cite{vavougios2018identification} used two-step clustering (similar to traditional distance based agglomerative clustering) to cluster based on pairwise differences at three time points for a single measure of functional disability, the Hoehn and Yahr scale. \cite{krishnagopal2020identifying} identified three clusters by maximizing community modularity through network-based trajectory profile clustering. Their methods relied on complete cases and involved simplifying disease severity variables into improving vs. non-improving at each timepoint. \cite{zhang2019data} used Long-Short Term Memory (LSTM) to represent each patient as a multi-dimensional time series and create a distance matrix for performing k-means for subgroup identification. \cite{krasniqidata} performed k-means and traditional distance based agglomerative clustering on dimensionally reduced regression coefficients of longitudinal measures, where separate longitudinal polynomial regressions were performed for each measure. \cite{de2019deep} introduced a variation autoencoder framework to identify three clusters using multivariate time series. Their methods integrated imputation into model training and only considered cluster results returning three clusters. These approaches are all reasonable, but they require either imputation or the use only complete cases only, inadequately capture temporal trends and correlations between variables, and/or use only baseline data to form clusters. 
 \newline

 Outside of studies in PD, methods have been developed to cluster curves/trajectories by shape or closeness. For example, \cite{genolini2016kmlshape} modified k-means to use the  Fr\'echet distance, capturing distance between curves. However, this approach is computationally complex, may place too much emphasis on noise within individual trajectories, and is designed to work with only one variable/curve. 
 \cite{luan2003clustering} demonstrated non-hierarchical clustering of time-course gene expression data using a mixed-effects model with basis splines and an expectation-maximization algorithm. \cite{heard2006quantitative} expanded their ideas to apply basis splines clustering in a Bayesian agglomerative clustering framework. However, both of these approaches allowed for only two dimensions (i.e. clusters of genes for multiple genes at multiple time points) but not three dimensions (e.g. clusters of genes for multiple genes per individual per time point or, in the application that follows, multiple individuals per variable per time point). Their methods are not designed to form clusters of individuals with multiple measures over time.  \newline

Another major concern with using longitudinal data for clustering and other applications is missing data. 
Imputation (generating values to serve as surrogates for the missing data) or list-wise deletion (excluding entire cases with any missing data) are the most common choices for most methods. List-wise deletion reduces sample size, sometimes drastically. According to \cite{de2019deep}, any imputation method generates some error, and these errors are even more impactful if clustering and imputation are executed separately.   \newline

Some temporal data represent time to an event and involve censoring. For these data, survival data analysis methods would be appropriate. Methods for clustering using time-to-event data have also been described in other literature. Many methods determine subgroups in time-to-event outcomes by applying post-hoc survival analyses to clusters pre-determined from static features.  If, however, the goal is to find subgroups exhibiting such differences, it is clearly preferable to include the time-to-event variables in the feature set used in the clustering algorithm. Methods that cluster a survival outcome typically use a mixture of Weibull distributions \citep{liverani2021clustering}. Similarly, \cite{proust2016joint} developed a joint model for longitudinal and Weibull time-to-event outcomes featuring latent classes and latent processes. However, in medical applications, semi-parametric methods (e.g. Cox proportional hazards model) and the non-parametric Kaplan-Meier estimator are far more common due to concerns that the Weibull distribution may not adequately fit the data. Other approaches to cluster with survival data rely on the Cox proportional hazards assumptions. In a method introduced by \cite{bair2004semi}, covariates highly related to survival via Cox proportional hazards models are selected and then clustered using a classical clustering approach. More flexible modeling frameworks may lead to better clustering results.  \cite{chapfuwa2020survival} developed a joint learning approach that optimizes latent representations of observations based on both covariates and time-to-event variables which are then clustered. All of these methods can only be applied with a single time-to-event variable and most require all other variables to be static. Uniquely, our methods allow for the use of multiple longitudinal and time-to-event variables to identify subgroups. \newline

We propose Divisive Hierarchical Bayesian Clustering for Longitudinal and Time-to-Event Data (D-HBC-LT) to address these concerns by forming clusters with both longitudinal and time-to-event data. This approach can appropriately capture the dependencies between variables measured on a single subject as well as temporal dependencies between repeated measurements on the same subject. D-HBC-LT is a divisive model-based method, starting with one cluster and splitting into smaller clusters at each step. It provides transparency in what metrics guide cluster membership while directly capturing differences in progression in the clustering process. In other words, while many clustering methods are complex and are seen as a ``black box" to many users (e.g., containing layers of transformations via an artificial neural network), our models rely on mixtures of very straightforward models and return the estimated parameters for each cluster. This provides clarity in terms of what variables are important for clustering and how they they differ between clusters. Further, this approach can avoid decreasing the sample size through list-wise deletion and the compounding errors of imputation. In modeling time-to-event data by Bayesian piecewise exponential (BPE) models, the shape of the survival curve is not restricted to a particular distribution, such as the Weibull, nor does it rely on the proportional hazards assumption between clusters. \newline

This paper is organized as follows. In Section 2, we describe the Divisive Hierarchical Bayesian Clustering (D-HBC) framework and longitudinal (LMM) and time-to-event (BPE) extensions to Divisive Hierarchical Bayesian Clustering  for Linear Mixed Models and Time-to-Event Data (D-HBC-LT). Section 3 describes the motivating data. Section 4 reports on the findings of the application of D-HBC to identifying subgroups in Parkinson's disease with longitudinal and survival data from the Parkinson's Progression Markers Initiative database. We provide concluding remarks in Section 5.

\section{Methods}
\label{sec:methods}

\subsection{Divisive Hierarchical Bayesian Clustering Algorithm}
Our proposed approach builds upon our recent work on divisive hierarchical Bayesian clustering (D-HBC) \red{\cite{post_this_on_arxiv}}. Briefly, D-HBC determines nested hard cluster assignments using overfitted finite mixture models, where clusters are defined as non-empty components of the mixture.  D-HBC does so without considering an exhaustive search of all possible splits by initializing each split with a rapid heuristic clustering approach (e.g. 2-means or 2-medoids), and proceeding with a coordinate ascent-like approach. This parametric hierarchical clustering algorithm increments at each step the Dirichlet concentration parameter $\alpha$ to move through the partition hierarchy to a higher number of clusters, and for each $\alpha$ the maximum a posteriori (MAP) estimates of cluster parameters and partitions is provided. 
Splits are continued until each observation is in its own cluster or some stopping criteria (e.g. a pre-specified maximum number of clusters is reached) are met.

\subsection{Divisive Hierarchical Bayesian Clustering for Multivariate Linear Mixed Models}

In applications for which the progression of disease or trajectories of data for each individual is of interest, data of repeated measures of multiple variables are often available. Frequently these variables are expected to be correlated with one another over time within individuals, so models that require conditional independence of each variable given the cluster assignment are based on an invalid assumption. A multivariate normal modeling approach may be more suitable by allowing correlation between variables within subjects. For this approach, it is recommended to restrict model matrices for fixed and random effects such that there is only one of each between all variables for each individual. In other words, some data points may be missing for some subjects, and the presented methods assume that the missingness pattern is missing completely at random or missing at random and that within an individual either all or no variables are missing at each time point. This may seem restrictive; however, in longitudinal studies it is common to measure variables of interest at the same or similar visit times. Thus, it is likely that this is a reasonable condition for many studies.  \newline

We begin by defining the multivariate normal modeling framework and then present a method by which parameters can be estimated. For individual $i$, $i=1,2\ldots,I$, $Y_i$ is a $n_i \times H$ data matrix, where columns of $Y_i$,  $Y_{i\cdot h}$, represent measurements of the $h^{th}$ variables over time and there are $H$  variables of interest. Rows of $Y_i$,  $Y_{it \cdot}$, represent measurements on each variable at the $t^{th}$ time point. For individual $i$, the fixed effects design matrix, $X_i$, is $n_i \times p$, and the random effects design matrix, $W_i$, is $n_i \times q$. If individual $i$ belongs to the $k^{th}$ cluster, the fixed effects coefficient matrix, $B_k$, is $p \times H$, and the matrix of random effects, $U_{ik}$, is $q \times H$. The error matrix is $n_i \times H$ and is denoted $E_{ik}$. For individual $i$ in cluster $k$, this model can be written as follows:
 \begin{align} \nonumber
        Y_i &= X_i B_k + W_i U_{ik} + E_{ik}\Leftrightarrow \\ \nonumber
        vec(Y_{ik}) & =
        vec(X_i B_k) + vec(W_i U_{ik}) + vec(E_{ik}) \\
        & =
        (I_H \otimes X_i) vec(B_k) + (I_H \otimes W_i) vec(U_{ik}) + vec(E_{ik}).
 \end{align} 
Cluster specific random effects for individual $i$, $U_{ik}$, are generated from a matrix normal distribution where the $q \times q$ covariance matrix corresponding to the rows (i.e., between random effects) is denoted as $G_k$, and the $H \times H$ covariance matrix corresponding to the columns (i.e., between variables) is the covariance matrix $\Omega_k$.  That is,
 \begin{align} \nonumber
        U_{ik} &\sim MN_{q,H}(0,G_k,\Omega_k) \Leftrightarrow \\
        vec(U_{ik}) & \sim N_{qH}(vec(0), \Omega_k \otimes G_k).
 \end{align} 
The error matrix is assumed to follow a matrix normal distribution where the $n_i\times n_i$ covariance among the rows (between observations) is $\Sigma_{ik}$, and the variance among the columns (between variables) is the $\Omega_k$,
 \begin{align} \nonumber
        E_{ik} &\sim MN_{n_i,H}(0,\Sigma_{ik},\Omega_k) \Leftrightarrow \\
        vec(E_{ik}) & \sim N_{n_iH}(vec(0), \Omega_k \otimes \Sigma_{ik}).
 \end{align}
Note that $\Omega_k$ is shared between the random effects and noise components of the model.  This greatly facilitates estimation, but also serves a reasonable assumption, namely that the way in which the variables relate to each other when measured cross-sectionally follows the same pattern as when these variables are measured over time. \newline 
 
Together, these distributional assumptions imply that the data for individual $i$ given the cluster assignment follows a matrix normal distribution, which can be expressed in vectorized notation as
 \begin{align} \nonumber
    & vec(Y_i|Z_{ik},\{B_k, G_k, \Sigma_{ik}, \Omega_k\}_{k=1}^K) &\\
    & \sim N_{n_i H}\bigg((I_H \otimes X_i) vec(B_k), (I_H \otimes W_i) (\Omega_k \otimes G_k) (I_H \otimes W_i)^T + \Omega_k \otimes \Sigma_{ik})\bigg)  
 \end{align}
or equivalently\\
\begin{align}
        Y_i &\sim MN_{n_i,H} \left(X_i B_k, W_iG_kW_i^T+\Sigma_{ik}, \Omega_k
\right). 
\end{align}

Using the matrix normal likelihood, the framework of D-HBC, and specifying priors on each component parameter, a posterior that supports correlations between variables can be constructed. It is convenient to work with a modified form of the posterior in order to fit longitudinal models and identify the MAP of the parameters more easily. Letting $\widetilde{Y}_i = Y_i \Omega_k^{-1/2}$ for subject $i$ belonging to cluster $k$, note that
 \begin{align}
    \widetilde{Y}_i &= X_i B_k \Omega_k^{-1/2} + W_i U_{ik} \Omega_k^{-1/2} + E_{ik} \Omega_k^{-1/2}.
 \end{align}
 Fitting a model to $\widetilde{Y}_i$ is easily done in vectorized form because
 \begin{align}
     vec(U_{ik}\Omega_k^{-1/2}) &\sim N_{qH}\bigg(vec(0), I_H \otimes G_k\bigg) \text{ and}\\
     vec(E_{ik}\Omega_k^{-1/2}) &\sim N_{n_i H}\bigg(vec(0), I_H \otimes \Sigma_{ik}\bigg).
 \end{align}
In other words, the columns of $\widetilde{Y}_i$ are independent. 
 \newline

Letting $\mathcal{I}_k := \{i: Z_{ik} = 1\}$ denote the set of all subjects in cluster $k$, and $ N_k := \sum_{i\in {\cal I}_k} n_i$ the total number of observations in cluster $k$, note that
\begin{align}
    Y_{\mathcal{I}_k} &\sim MN_{N_k,H} \left(
X_{\mathcal{I}_k} B_k, W_{\mathcal{I}_k} G_k W_{\mathcal{I}_k}^T +  \Sigma_k, \Omega_k
\right),
\end{align}
where the $N_k\times H$ matrix $Y_{\mathcal{I}_k}$ is obtained from stacking $\{Y_i\}_{i\in\mathcal{I}_k}$, and similarly for the $N_k \times p$ model matrix for fixed effects, $X_{\mathcal{I}_k}$, and the $N_k \times q$ model matrix for random effects, $W_{\mathcal{I}_k}$, and where $\Sigma_k$ is a $N_k \times N_k$ block diagonal covariance matrix whose blocks correspond to $\Sigma_{ik}$'s. 
\newline

All cluster parameters can be estimated in a coordinate ascent-like approach, iterating between finding parameters using $\widetilde{Y}_{\mathcal{I}_k}$ given $\Omega_k$ (i.e. $B_k$, $G_k$, and $\Sigma_k$) and updating the $\Omega_k$ given $Y_{\mathcal{I}_k}$ and all other parameters. A semi-conjugate inverse Wishart prior, $\Omega_k \sim IW(\Psi,\nu)$, leads to a closed form update for $\Omega_k$,
\begin{align*}
	\widehat{\Omega_k} &= \frac{\widehat{\Psi}}{\hat{\nu} + H + 1},
\end{align*}
where $\widehat\Psi = \Psi + (Y_{\mathcal{I}_k} - X_{\mathcal{I}_k} B_k)^T(W_{\mathcal{I}_k} G_k W_{\mathcal{I}_k}^T + \Sigma_k)^{-1} (Y_{\mathcal{I}_k} - X_{\mathcal{I}_k} B_k)$ and $\hat{\nu} = \nu + n_i$.

To reduce the computational burden of inverting the $N_k \times N_k$ matrix $W_{\mathcal{I}_k} G_k W_{\mathcal{I}_k}^T + \Sigma_k$, the Woodbury matrix identity can be used:
\begin{align*}
    (W_{\mathcal{I}_k} G_k W_{\mathcal{I}_k}^T + \Sigma_k)^{-1} &=
    \Sigma_k^{-1} - \Sigma_k^{-1} W_{\mathcal{I}_k} (G_k^{-1} + W_{\mathcal{I}_k}^T\Sigma_k^{-1} W_{\mathcal{I}_k})^{-1}W_{\mathcal{I}_k}^T \Sigma_k^{-1}.
\end{align*}
Since $\Sigma_k$ is a block diagonal matrix, this inversion is not onerous.  In fact, one often uses $\Sigma_k = \sigma^2 I_{N_k}$ (differences in scale between variables is captured in $\Omega_k$), i.e., the dependence between observations is captured through the random effects, which further reduces the complexity of this quantity by making the largest matrix to invert of dimension $q \times q$. Given $\Omega_k$, obtaining the solutions for updating the remaining parameters is straightforward, based on traditional Bayesian linear mixed model (LMM) using updated $\tilde{Y}_{\mathcal{I}_k}$.

\subsubsection{Bayesian Piecewise Exponential Model Overview}
The Bayesian piecewise exponential (BPE) model, also known as the piecewise constant hazard model, is a semi-parametric survival model known for being simple yet flexible, accommodating many shapes of survival data. \cite{ibrahim2001bayesian} provides a background of and examples using BPE. One can begin by considering a single time-to-event variable, although more than one time-to-event variable may be included in D-HBC-LT under the conditional independence assumption. The data are denoted $\{t_i\}_{i=1}^I$ for event/censoring times and $\{d_i\}_{i=1}^I$ for indicators of event occurrence with subscript $i$ representing individual subjects. As is typical with survival data, some observations may be right censored, in which case $d_i = 0$. Typically, BPE is specified with the baseline hazard as piecewise constant; however, because covariates will not be included, all hazards can be assumed to each be piecewise constant, $\lambda(t) = \lambda_j$ for all $t \in [a_{j-1},a_j)$, where $0 = a_0 < a_1 < \dots < a_J$ are the changepoints in the hazard function. These piecewise constant hazards provide a continuous survival curve. For an event in the $j^{th}$ interval, survival is given by
\begin{align*}
        S(t) &= exp\bigg\{\int_0^t \lambda(u)du \bigg\}\\
        &= exp\{\lambda_1 (a_1 - a_0) + \lambda_2 (a_2 - a_1) + \dots +
        \lambda_j (t - a_j) \}.
    \end{align*}
The BPE model is convenient because it can be written as an equivalent Poisson regression model \citep{holford1980analysis, laird1981covariance}. The derivation as it relates to the presented methods and application is provided in the Supplemental Materials for convenience. In the analysis that follows, a fixed number of changepoints, or grid intervals, that are independent of the data are selected, without having so many intervals as to make the model non-parameteric. This approach is supported by \cite{ibrahim2001bayesian}. 

\subsubsection{Divisive Hierarchical Bayesian Clustering for Longitudinal Data and Time-to-Event Data (D-HBC-LT)}

Building upon the multivariate LMM, BPE, and D-HBC described above, D-HBC-LT will be outlined to identify clusters based on temporal data for each individual. The classification matrix is defined $Z = \big[Z_{ik}\big]$, $i=1,\ldots,I$, $k=1,\ldots,K$: if individual $i$ belongs to cluster $k$, $k=1,2,\ldots,K$, then $Z_{ik}$ equals one, and zero otherwise. Cluster assignments are assumed to be static and do not change with time. Typically $K$, the number of mixture components, is set to be $I$, and a cluster is defined as a non-empty component. Thus the mixture model is overfitted and for many configurations along the hierarchy, most components will be empty. Starting with the classification likelihood
\begin{align}
    \pi(Y_1,\ldots,Y_I|Z, \theta) &= \prod_{k=1}^K \prod_{i=1}^I  [f(Y_i|\theta_k)]^{Z_{ik}},
\end{align}
where $f(\cdot|\theta_k)$ is the likelihood function for an individual, parameterized by 
\begin{align*}
\theta_k:= \{B_k, G_k, \Sigma_k, \lambda_{1k}, \dots, \lambda_{Jk}\}.
\end{align*}
Assuming conditional independence of the longitudinal variable set and each of the time-to-event variables given the cluster assignments, $f$ is a product of the likelihood of the LMM of the $k^{th}$ cluster given by
\begin{align}
     Y_{\mathcal{I}_k} &\sim MN_{N_k,H} \left(
X_{\mathcal{I}_k} B_k, W_{\mathcal{I}_k} G_k W_{\mathcal{I}_k}^T + \Sigma_k, \Omega_k \right),
\end{align}
and the BPE model given by
\begin{align}
    \pi(y_{\mathcal{I}_k}|\lambda_{\cdot k})
    &= \prod_i^I \prod_j^J \bigg[(\lambda_{jk} T_{ijk} )^{N_{ijk}}e^{\lambda_{jk} T_{ijk}}\bigg],
\end{align} 
where $N_{ijk}$ is an indicator of failure for subject $i$ from cluster $k$ occurring in the $j^{th}$ interval and $T_{ijk}$ is the subject's risk time spent within the interval, i.e.,
\begin{align*}
    N_{ijk} &= \mathbf{1}_{[t_i \in (a_{j-1},a_j) \text{ and } d_i = 1]}\\
    T_{ijk} &= \begin{cases}
    min(t_i, a_j) - a_{j-1} & t_i  > a_{j-1}\\
     0 & t_i  < a_{j-1}.
    \end{cases}
\end{align*}

As with D-HBC, the prior on $\theta_k$ will depend on whether or not the component is empty. 
Specifically, the prior on the component parameters is set to have the following form:
\begin{align}
    \pi(\theta_k|Z) &= \begin{cases}
    \pi_1(\theta)
    & \sum_i Z_{ik} > 0\\
    {\pi_{0}(\theta) = \frac{1_{[\theta \in {\cal B}]}}{\int_{{\cal B}} d{\theta}}
      }
    & \sum_i Z_{ik} = 0.
    \end{cases}
\end{align}
Here, $\pi_1(\theta_k)$ is the prior density function given that cluster $k$ is non-empty (i.e. contains at least one observation). For an empty cluster, $\pi_0(\theta_k)$ is set to a uniform prior over ${\cal B}:= \widetilde{{\cal B}}\cap \Theta$, where $\widetilde{{\cal B}}$ is some ball around an interior point of the parameter space, $\Theta$. For more on this conditional prior over the component parameters, see Chapter 2. In the analyses that follow, $\pi_0(\theta_k) = 1$.  \newline

A Dirichlet-multinomial prior is placed on the component weights, $\xi:=(\xi_1,\ldots,\xi_K)$, and the classifiers, and a uniform prior on the Dirichlet shape parameter, $\alpha$. We integrate out the component weights, which are nuisance parameters, such that the log posterior is
\begin{align}\nonumber
     &\log[\pi(\{\theta_k\}_{k=1}^K, Z | Y)] &\\
     &= const + \log[\pi(Y|\{\theta_k\}_{k=1}^K, Z)] + \log[\pi(\{\theta_k\}_{k=1}^K|Z)] + \log\bigg[\frac{\Gamma(K\alpha)}{ [\Gamma(\alpha)]^K} \frac{\prod_{k=1}^K \Gamma(n_k + \alpha)}{\Gamma(N + K\alpha)}\bigg], 
\end{align}
where $n_k = \sum_i Z_{ik}$ is the number of individuals assigned to cluster $k$.\newline

\subsubsection{D-HBC-LT Algorithm}

The D-HBC-LT algorithm is similar to the previously described D-HBC algorithm. The goal of this algorithm is to identify a~posterior mode for each number of clusters, constrained to provide a nested hierarchical structure. Starting with all observations in one cluster, splits that lead to the greatest increase in the posterior are successively selected. There are some important differences between D-HBC for static data and D-HBC-LT for temporal data, which will be elucidated in what follows.  \newline

Mixed effects models depend on having more than one group for random effects to be identifiable (groups are defined as individuals $i$ here), so forming clusters with LMM requires that each cluster contain more than one individual. This means that at the formation of new clusters, each candidate split must contain more than one observation or the candidate split will not be selected. Additionally, this means that the construction of a ``complete" dendrogram, from $1$ to $N$ is not possible. Fortunately, the number of useful clusters tends to be much closer to $1$ than to $N$.\newline

To address the restrictions with LMM, a split is not attempted if the number of subjects is less than double the dimension of the cluster parameter vector for LMM. Next, if either of the initializing candidate clusters are smaller than the dimension of the cluster parameter vector, the number of subjects needed to make the smaller candidate cluster the size of the dimension of the cluster parameter vector are moved to the smaller candidate cluster from the larger candidate cluster. This shift of subjects at initialization is performed by moving those subjects with a weaker preference for the larger candidate cluster. For example, if using 2-medoids with a selected distance metric (e.g. a distance between clustering variable slopes of subjects), one would move those subjects with the smallest difference between the distance to the larger candidate cluster medoid and the smaller candidate cluster medoid. After initialization, if at any point the coordinate ascent between candidate cluster parameters and candidate cluster assignments to split cluster $k$ selects fewer subjects than the dimension of the cluster parameter vector for LMM of either candidate cluster, a split of cluster $k$ will not be recommended. This is because the evidence is not strong enough to support a split of cluster $k$ with the current model. \newline

The D-HBC-LT algorithm is outlined below. 

\include{Paper2/long_algortihm}

\section{Data Structure and Background}
\label{sec:background}
Longitudinal data from subjects recently diagnosed with idiopathic PD were downloaded from the Parkinson's Progression Markers Initiative (PPMI) database on March 15, 2021. Enrolled participants were required to meet several criteria: a) be over 30 years old, b) have a low level of functional disability as demonstrated by a Hoehn and Yahr (H\&Y) stage of 1 or 2, c) have either an asymmetric resting tremor or asymmetric bradykinesia or two of the following symptoms: bradykinesia, resting tremor, and rigidity with recent diagnosis of PD, and d) are not yet treated by PD medications of symptomatic therapy \citep{zhang2019data}. This cohort was intended to be early in their course of PD, so that progression could be more easily identified.  \newline

Variables for clustering individuals were selected to be common and meaningful measures of disease severity in PD. The longitudinal variables used for clustering included: the Movement Disorder Society Unified Parkinson's Disease Rating Scale (MDS-UPDRS) Parts 1-3, the Epworth Sleepiness Scale (ESS), the Symbol Digit Modalities Test (SDMT), and Letter-Number Sequencing (LNS). The MDS-UPDRS was developed as a comprehensive assessment of several domains of PD \citep{goetz2008movement}. The MDS-UPDRS is recognized as ``a benchmark for PD assessment" 
used to monitor the motor and non-motor function of Parkinson's disease patients over time \citep{bhidayasiri2017clinical}. MDS-UPDRS 1 is Non-Motor Aspects of
Experiences of Daily Living, assessing cognition, behavior, and mood. MDS-UPDRS 2 is Motor Aspects of Experiences of Daily Living, querying patients on skills fundamental to self care and independence, such as speech, writing, eating, and dressing. MDS-UPDRS 3 is the Motor Examination, obtained by having a trained clinician perform 33 assessments based on 18 items (many are performed on left and right) of motor function \citep{goetz2008movement}. With each MDS-UPDRS, higher scores indicate more severe disease states. The ESS is a brief self-administered questionnaire ($<$5 min) to assess daytime sleepiness \citep{hagell2007measurement, kumar2003excessive}. Fatigue and excessive daytime sleepiness is a common feature of PD, associated with more advanced PD and having many potential causes (drug therapy via higher levodopa-equivalent doses and dopamine agonists; sleep disorders such as insomnia or restless legs syndrome; disruption of circadian rhythm) \citep{arnulf2005excessive, hagell2007measurement, kumar2003excessive}. The ESS ranges from 0 to 24, with values greater than 10 typically recognized as excessive somnolence \cite{arnulf2005excessive}. The SDMT is a brief test to assess cognition and processing speeds \citep{smith1973symbol, grolez2020functional}. It is estimated that 80\% of PD patients develop cognitive impairment during the course of their disease \citep{goldman2018cognitive}. Lower scores on the SDMT are associated with slower processing speeds \citep{grolez2020functional}. The LNS is a short assessment of working memory, particularly the ability to process and resequence information \citep{mielicki2018measuring}. It is important to note that higher severity in scores within the parts of the MDS-UPDRS and between MDS-UPDRS parts and ESS have been demonstrated to be correlated \citep{goetz2008movement, kumar2003excessive}. Additionally, studies have demonstrated an age effect on each scale \citep{goetz2008movement, hagell2007measurement, smith1973symbol, pezzuti2017number}. \newline

The time-to-event variable was time from PD diagnosis to initiation of symptomatic pharmacological therapy. PD symptomatic therapies are not disease modifying; they alleviate or decrease functional disability from motor symptoms but they do not alter the disease progression rate. The decision to initiate symptomatic medical therapy for PD is influenced by several factors; however, one key factor is the extent to which motor symptoms disrupt a patient's quality of life \citep{connolly2014pharmacological}. In other words, choosing to initiate PD medications represents a decision that alleviating effects of the medications likely outweigh the risk of adverse effects of the therapy \citep{connolly2014pharmacological}. Thus initiating symptomatic therapy indicates progression to the point that symptoms are no longer acceptable.  \newline

In order to ensure that our clustering results did not simply reflect differences in age, prior to clustering each longitudinal variable vector was projected onto the space orthogonal to the column space of the fixed effects design matrix using only age at baseline as a covariate. For our time to initiation of symptomatic pharmacological therapy, BPE models featured intervals with changepoints for every 6 months from diagnosis adjusted such that no changepoint was equal to any subjects event or censoring time, as the model requires. Sensitivity analyses were performed using 1 year intervals for BPE and fitting survival data to the Weibull distribution. Clustering results were robust to these changes in the time-to-event analysis but sensitive to data driven changepoints based on time on study quantiles. \newline 

There were 355 subjects with data on all clustering variables, with longitudinal variables needing to have the same missingness pattern within an individual (e.g. subject 1 must have completed MDS-UPDRS 1, 2, and 3, ESS, LNS, and SDMT at year 1 for these data to be included, but subject 1 could have data at only years 1, 2, and 4; see previous chapter for more information about the model matrix for correlated longitudinal variables). To avoid any biases due to attrition (due to dropout or because some scheduled visits will occur in the future), only measurements up to 5 years after diagnosis were included for all variables. The model presented features linear time and a random intercept. Cluster analyses were implemented in R in conjunction with the blme package \citep{chung}.

\section{Parkinson's Disease Subtypes}
\label{sec:parkinsons}
\begin{figure}[H]
    \centering
    \includegraphics[scale = .23]{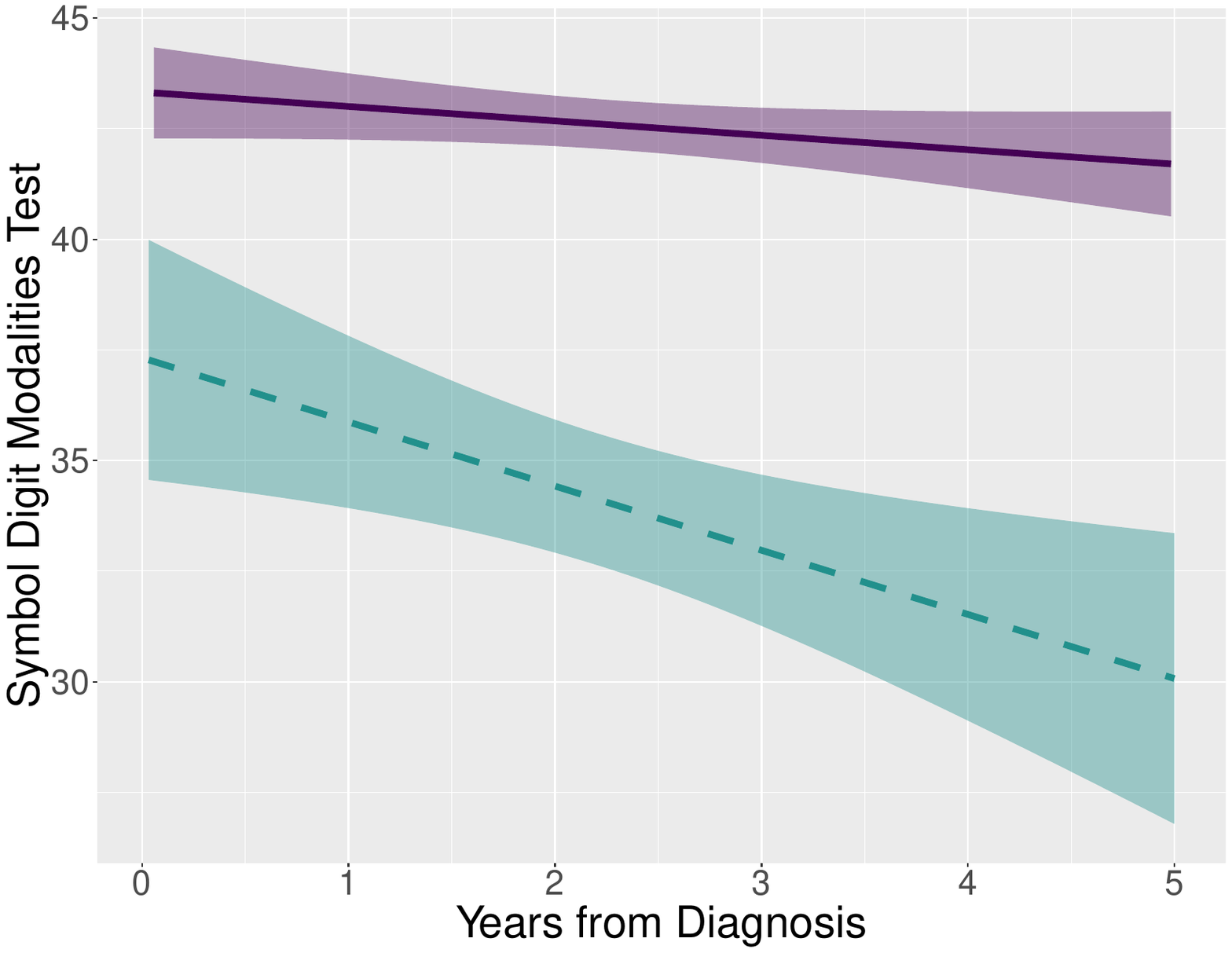}
    \includegraphics[scale = .23]{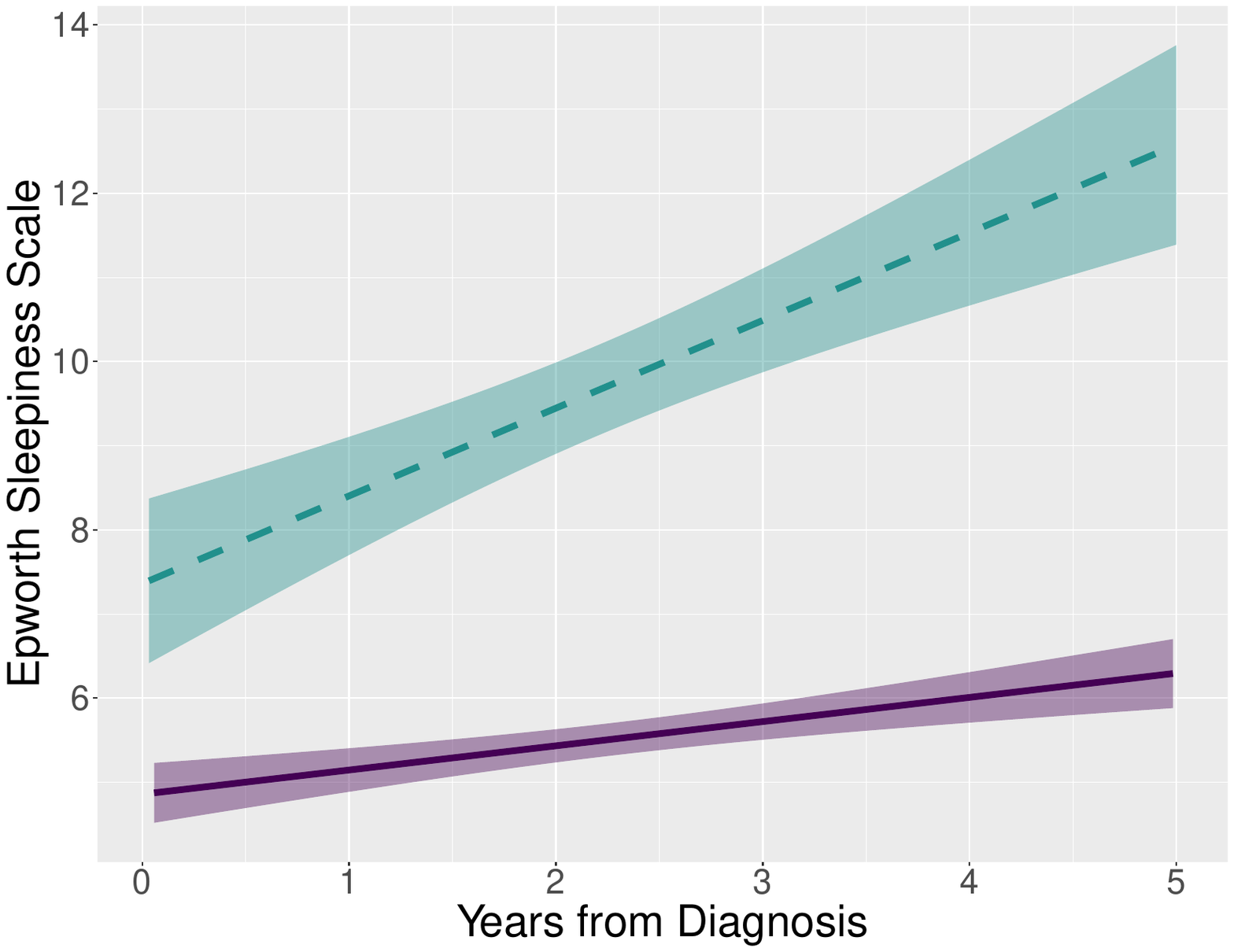}\\
    \includegraphics[scale = .23]{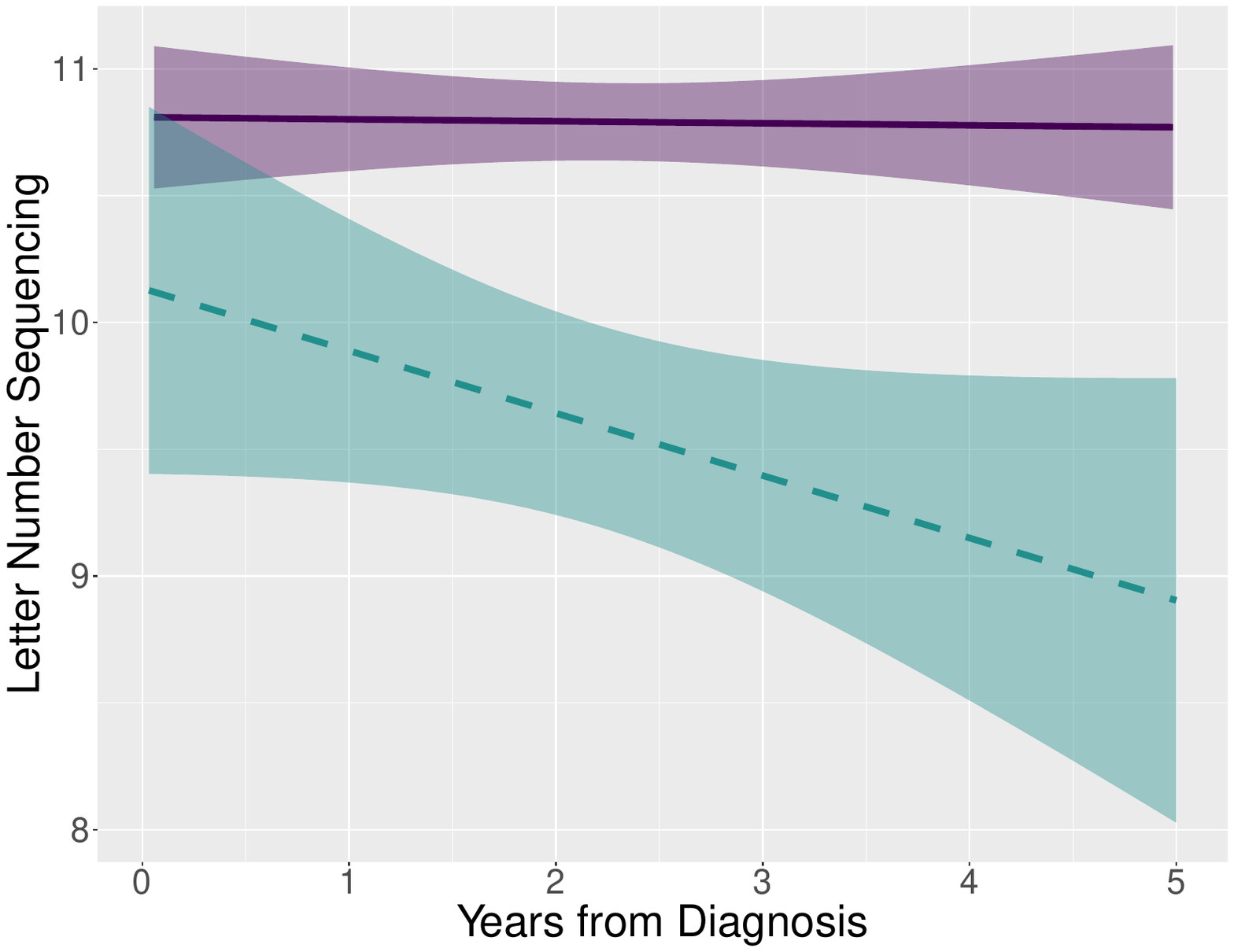}
    \includegraphics[scale = .23]{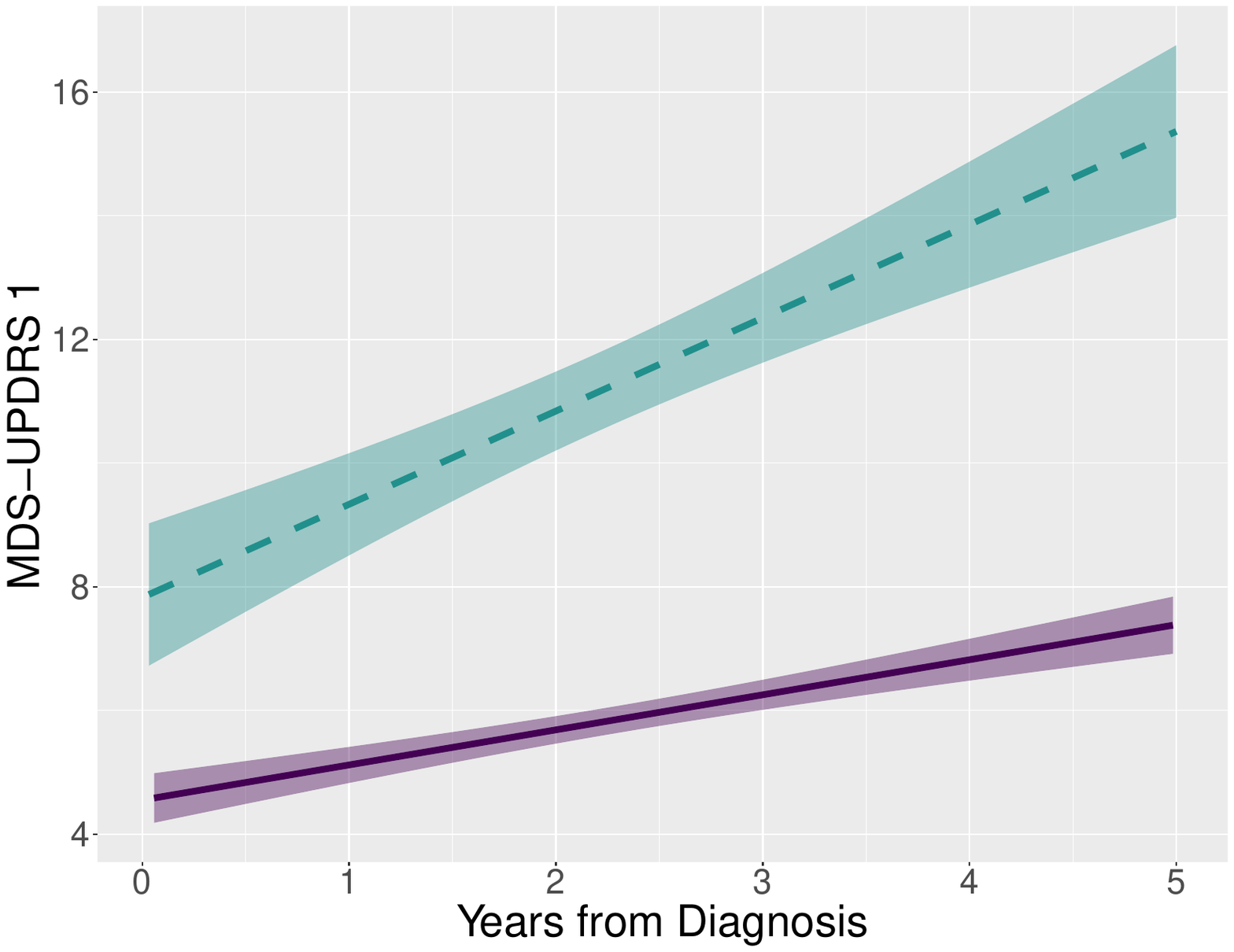}\\
    \includegraphics[scale = .23]{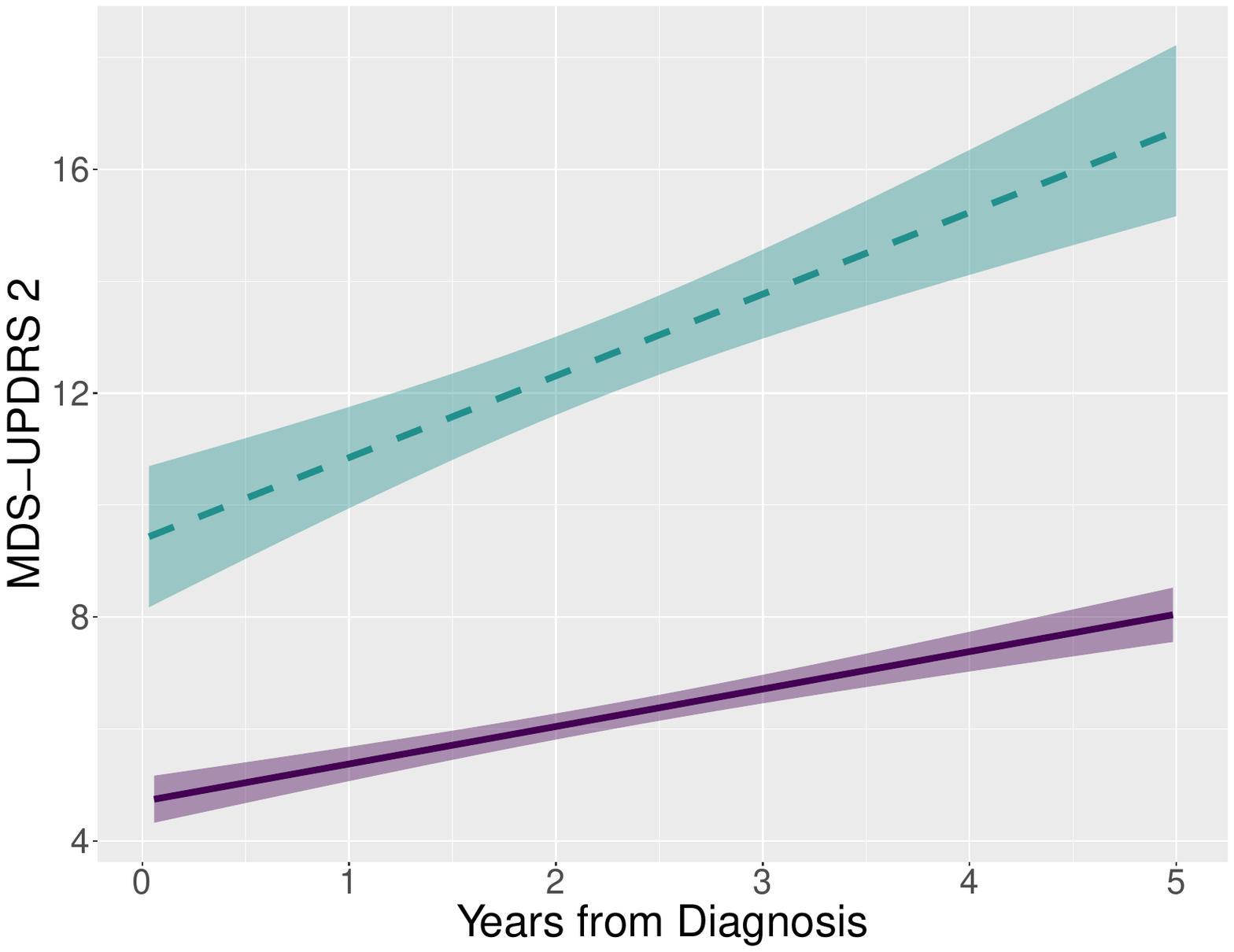}
    \includegraphics[scale = .23]{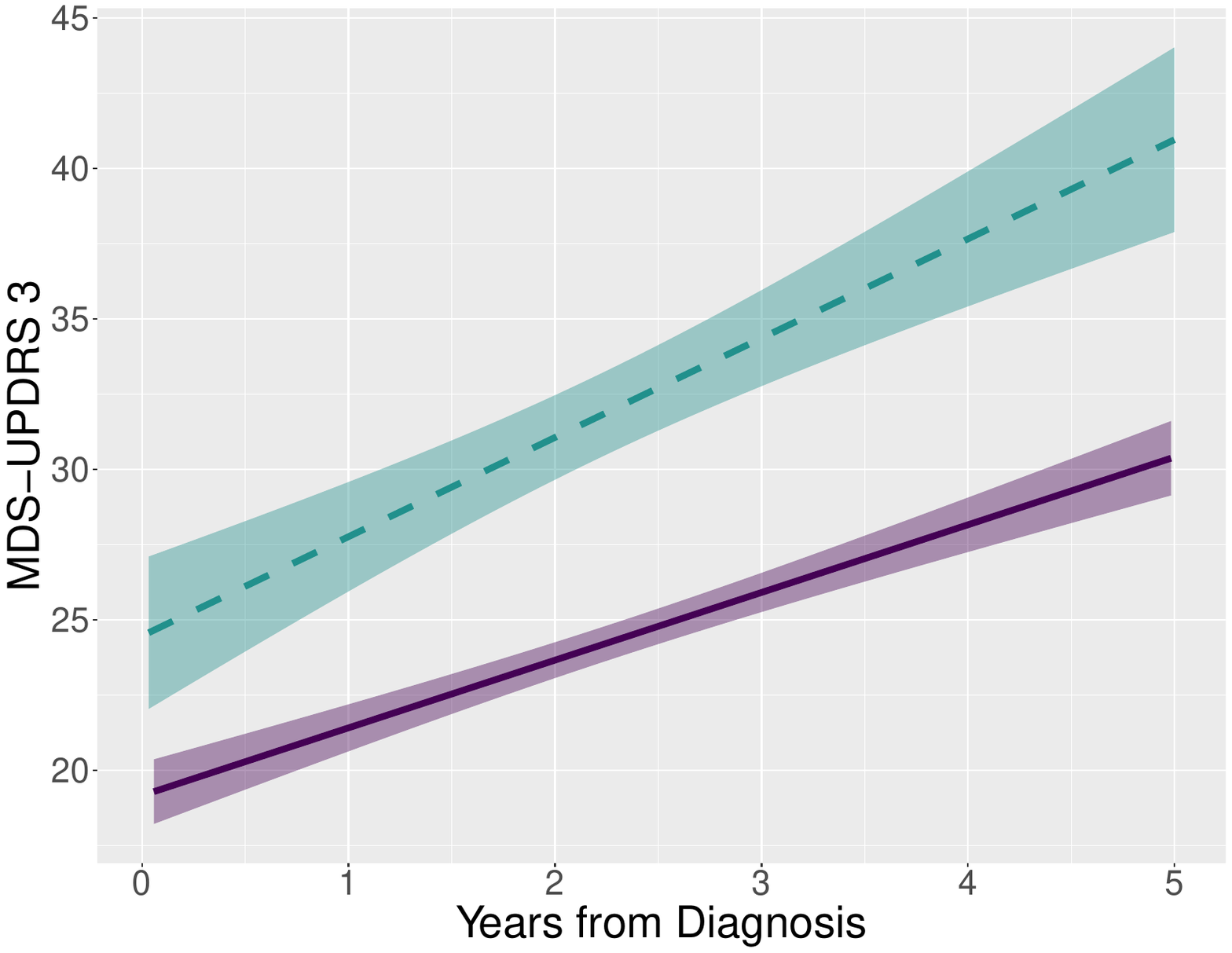}\\
    \caption{Two clusters demonstrated by mean trajectory and 95\% CI of clustering variables: SDMT (top left), ESS (top right), LNS (middle left), MDS-UPDRS 1 (middle right), MDS-UPDRS 2 (bottom), MDS-UPDRS 3 (bottom right).}
    \label{fig:spagclus}
\end{figure}

\begin{figure}[H]
    \centering
    \includegraphics[scale = .3]{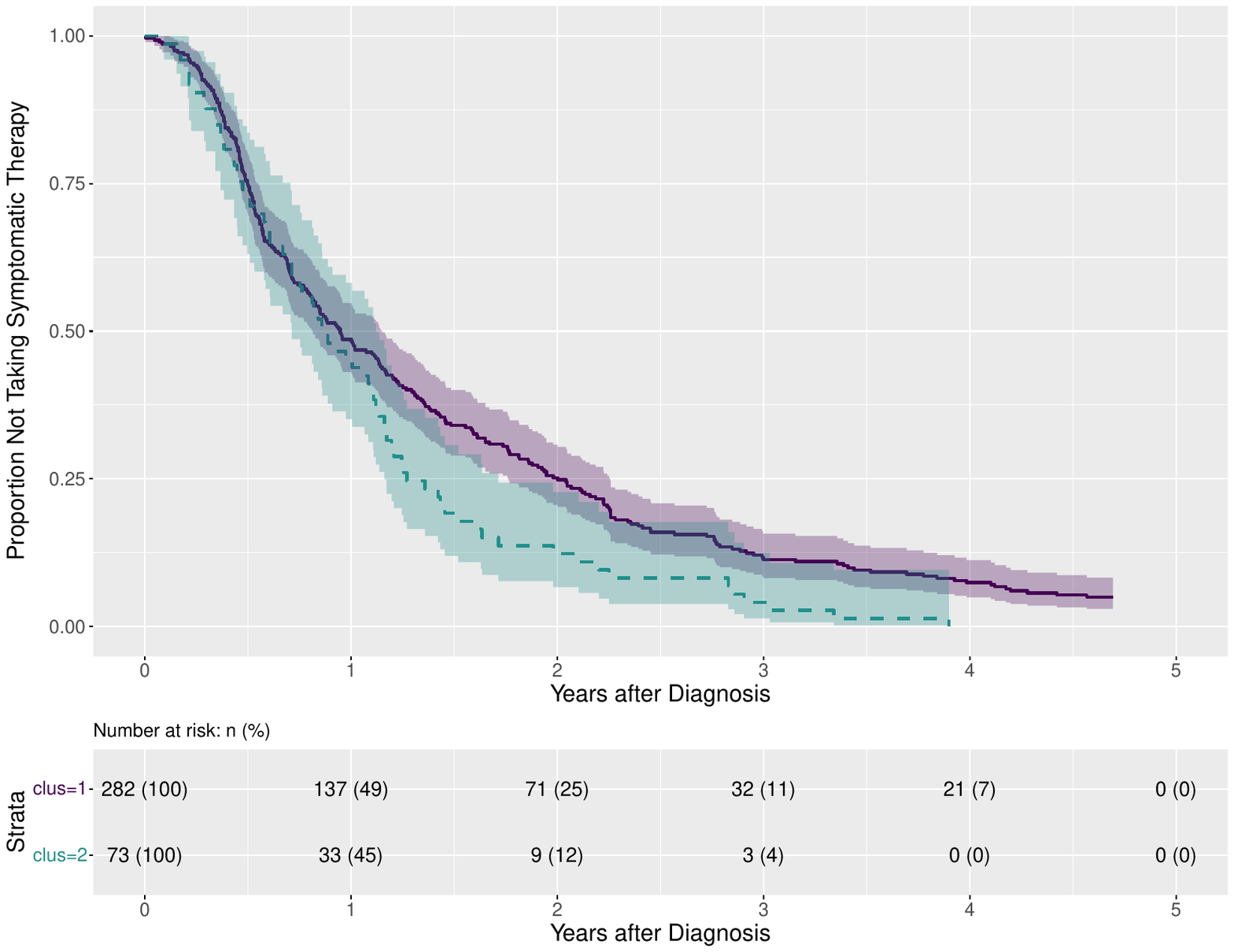}\\
    \caption{Two clusters demonstrated by Kaplan-Meier plot for time to symptomatic therapy initiation.}
    \label{fig:survclus}
\end{figure}

D-HBC-LT identified hierarchical clusters. Based on the consideration of a reasonable cluster size and the extent of the overlap/separation among the clusters, as well as cluster membership stability (discussed later in this section), we selected two clusters. These clusters are represented by the mean trajectory and 95\% confidence intervals for each longitudinal clustering variable in Figure \ref{fig:spagclus} and by a Kaplan-Meier plot in Figure \ref{fig:survclus}. A summary of coefficients returned by the Bayesian LMM of clustering variables (projected onto the space orthogonal to the column space of the fixed effects design matrix using only age at baseline as a covariate) for two clusters are presented in in \ref{tab:coef}. The mean linear trajectory plots with 95\% confidence intervals demonstrate steeper slopes in the dashed green cluster versus the solid purple cluster, each in the direction of decreased function or performance. In the Kaplan-Meier plot, approximately 1 year following diagnosis, those in the dashed green cluster tended to require symptomatic therapy sooner than those in the solid purple cluster. 

\begin{table}[H]
    \centering
    \begin{tabular}{|c|c|c|c|}
    \hline
         \multicolumn{2}{|c}{Clustering Variable} & Solid Purple & Dashed Green  \\
         \hline
         \multirow{2}{*}{SDMT} & Intercept &  0.2361 & -0.3531\\
          & Slope & -0.07986 & -0.4423 \\
          \hline
         \multirow{2}{*}{ESS} & Intercept & -0.3777 & 0.26362 \\
          & Slope &   0.2048 & 0.7016\\
          \hline
        \multirow{2}{*}{LNS}  & Intercept & 0.08410 &  -0.1439  \\
           & Slope & 0.004659 &  -0.2903 \\
         \hline
         \multirow{2}{*}{UPDRS 1} & Intercept &  -0.5072 & 0.1977 \\
           & Slope &  0.3368 & 0.8681\\
           \hline
         \multirow{2}{*}{UPDRS 2} & Intercept &  -0.5497 & 0.3715  \\
           & Slope & 0.3619 & 0.7616\\
           \hline
         \multirow{2}{*}{UPDRS 3} & Intercept & -0.6037  & -0.1254 \\
            & Slope & 0.5567 & 0.8039   \\
            \hline
    \end{tabular}
    \caption{Regression Coefficients of Linear Mixed Effects Model of Variable Projections for Two Clusters}
    \label{tab:coef}
\end{table}

\begin{figure}[H]
    \centering
        \includegraphics[scale = .23]{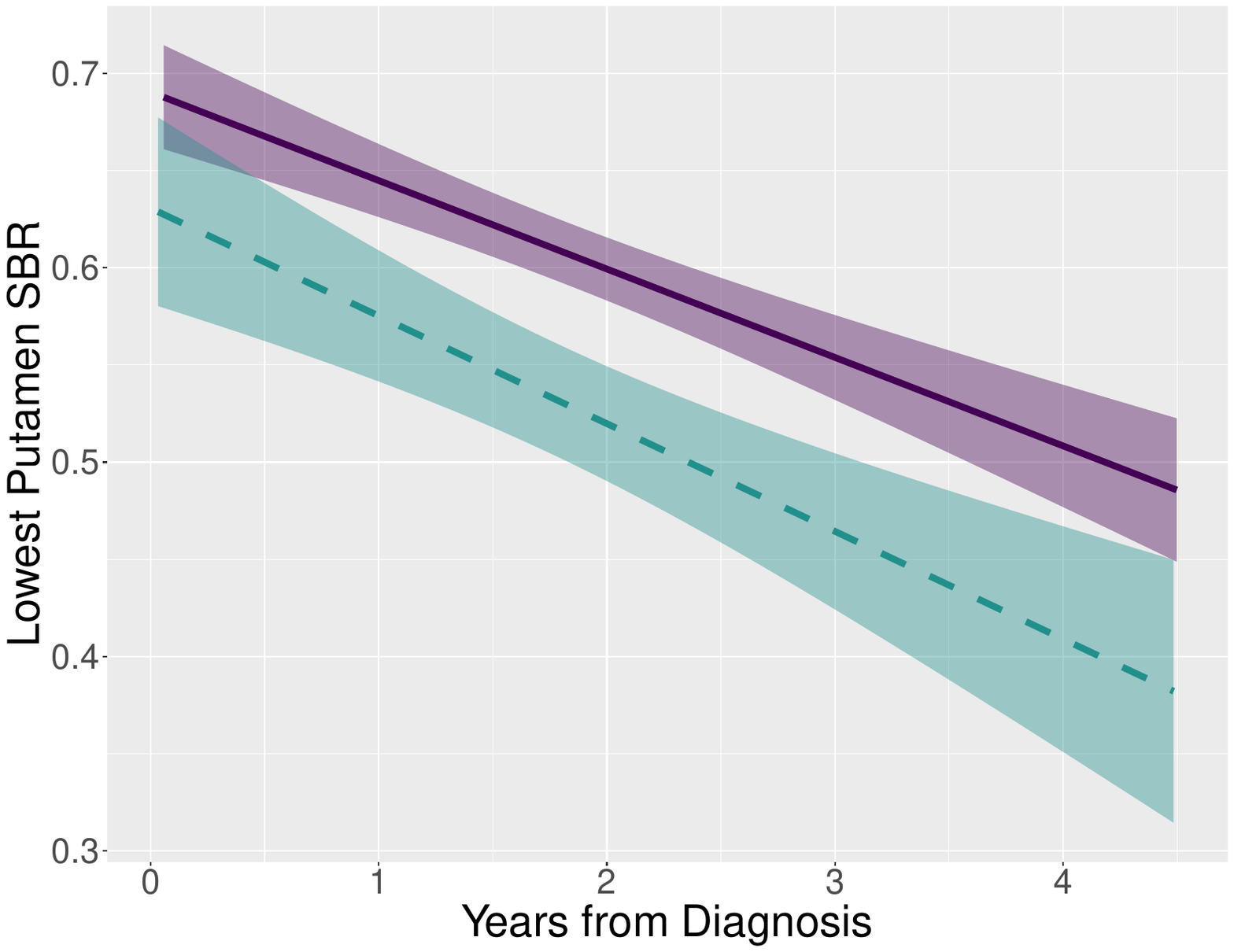}
    \includegraphics[scale = .23]{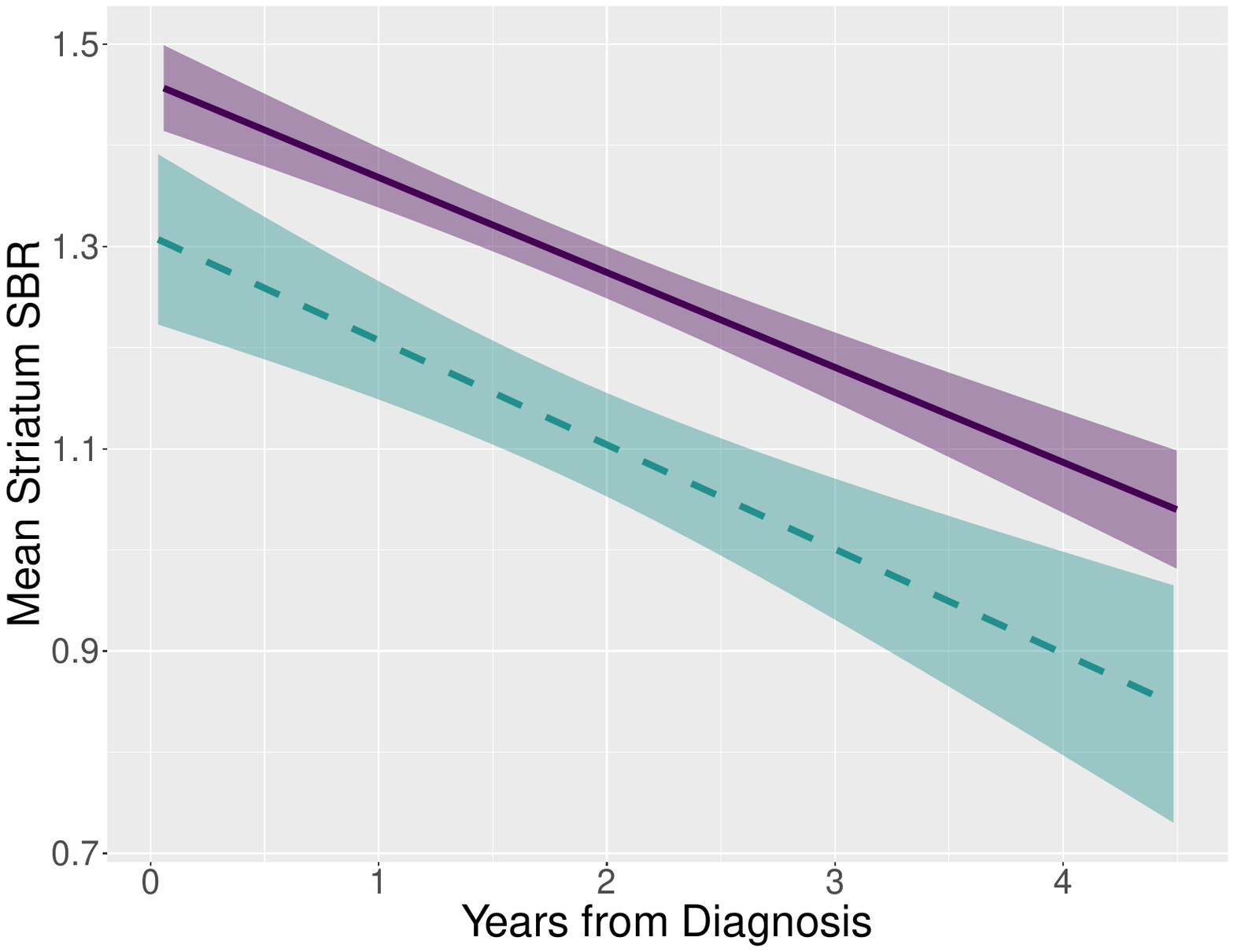}\\
    \includegraphics[scale = .23]{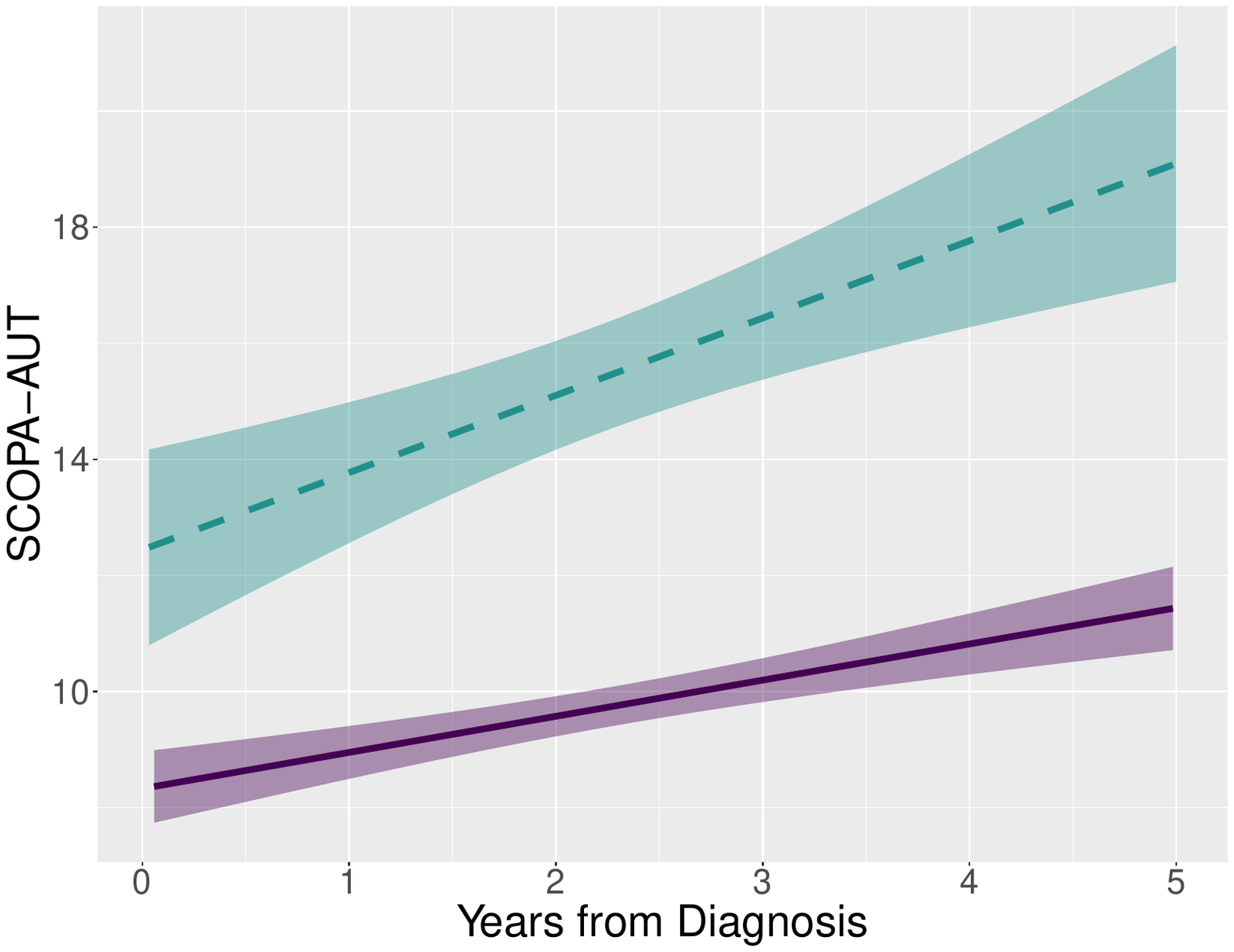} 
    \includegraphics[scale = .23]{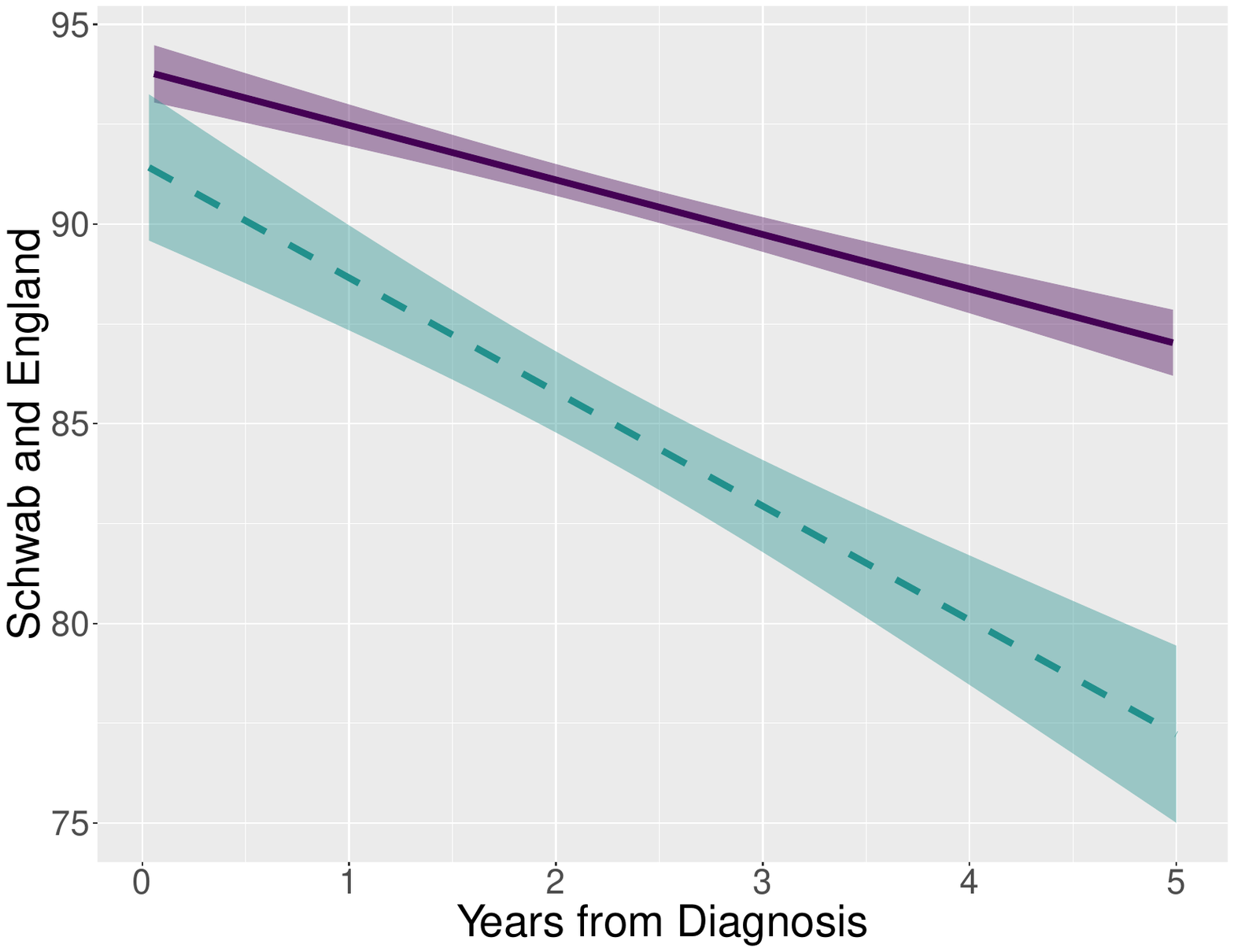} \\
    \includegraphics[scale = .23]{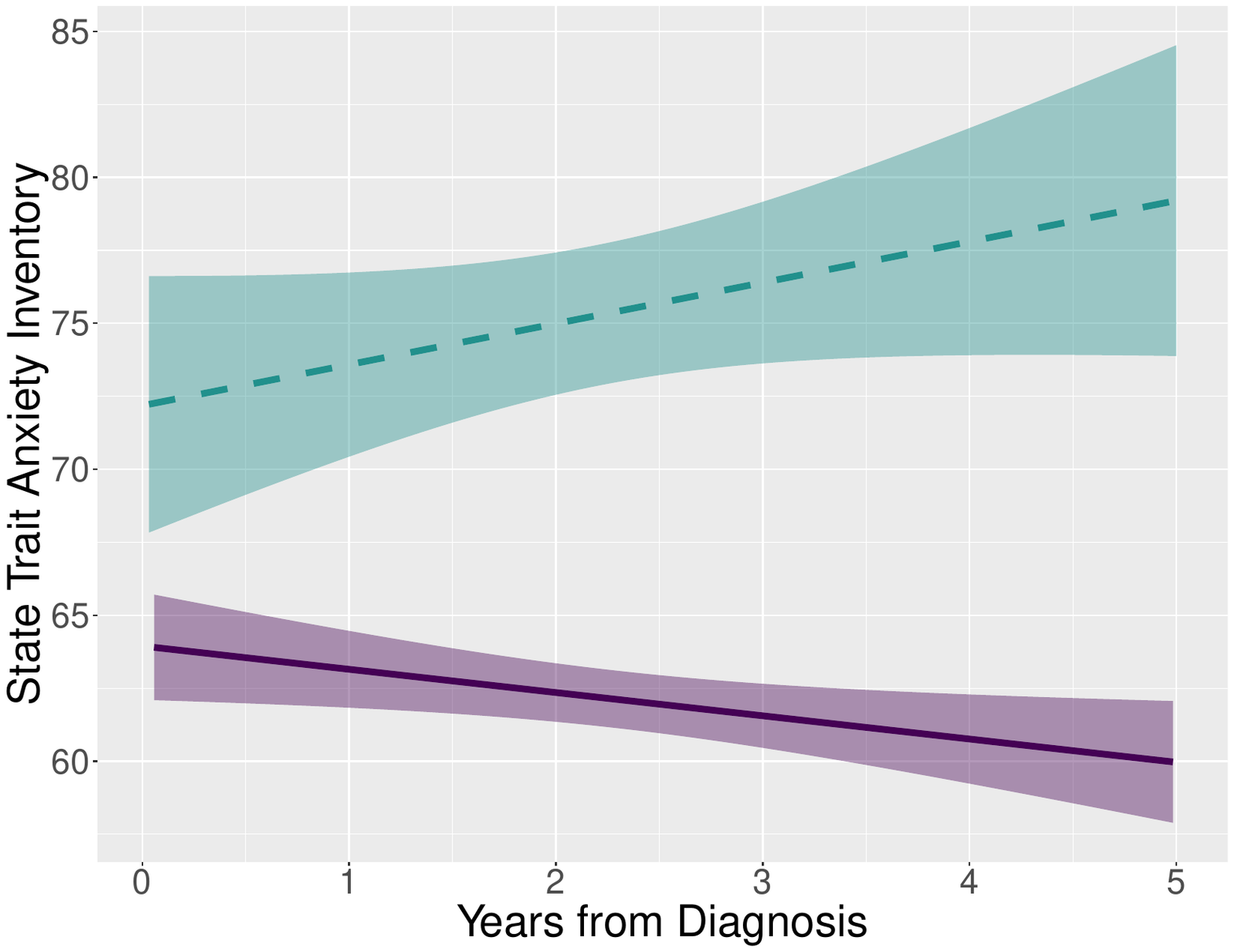}
    \includegraphics[scale = .23]{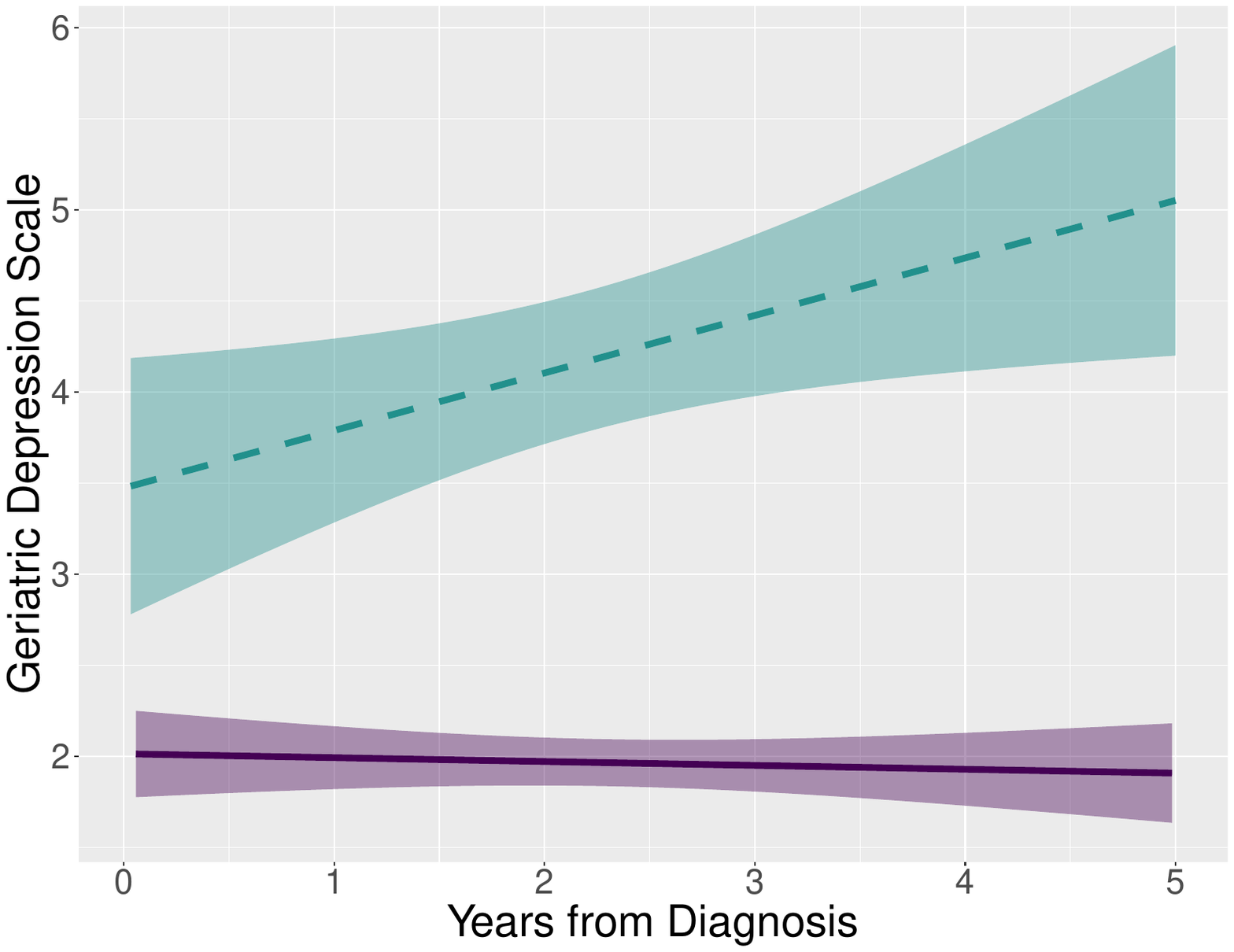}\\
    \includegraphics[scale = .23]{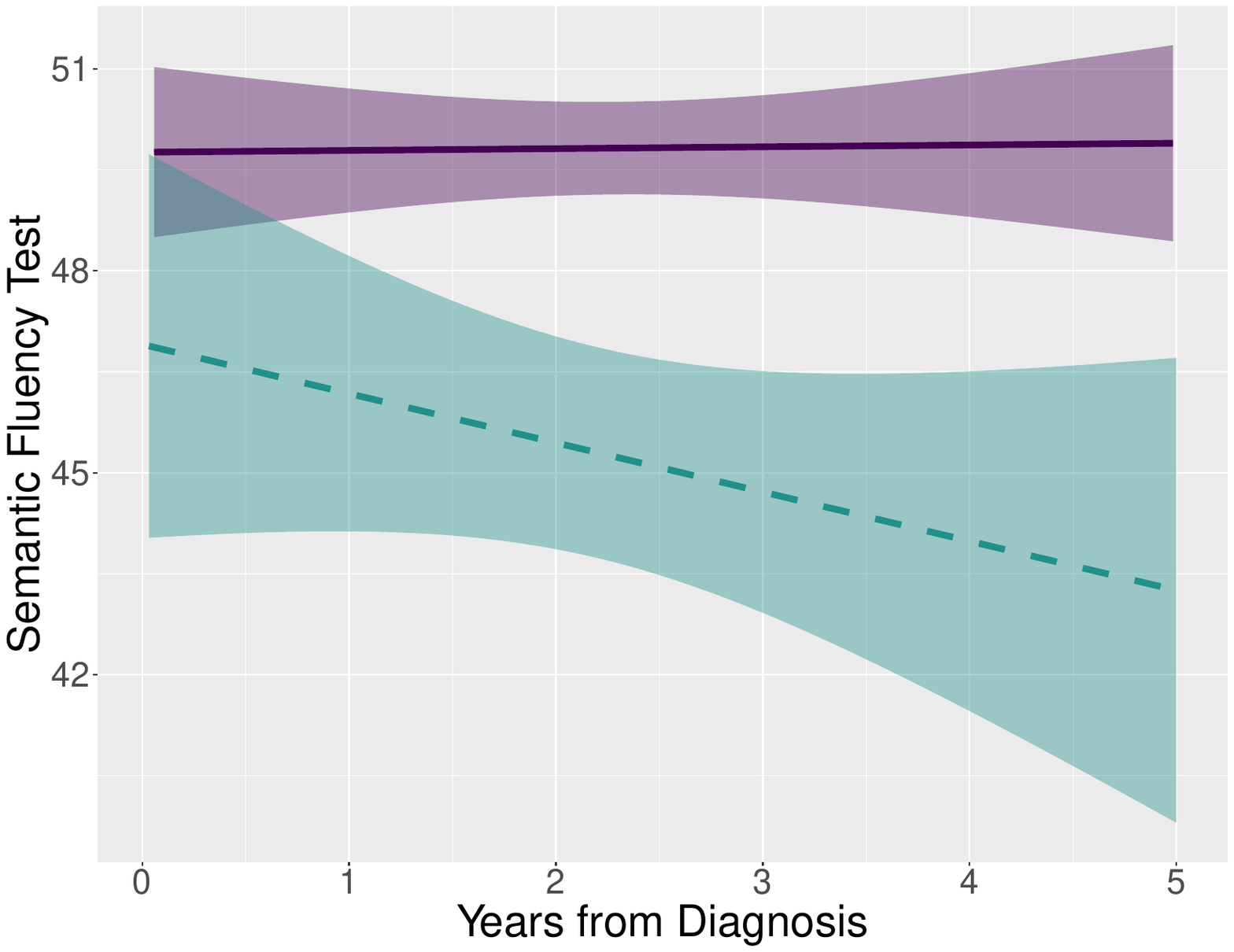}
    \includegraphics[scale = .23]{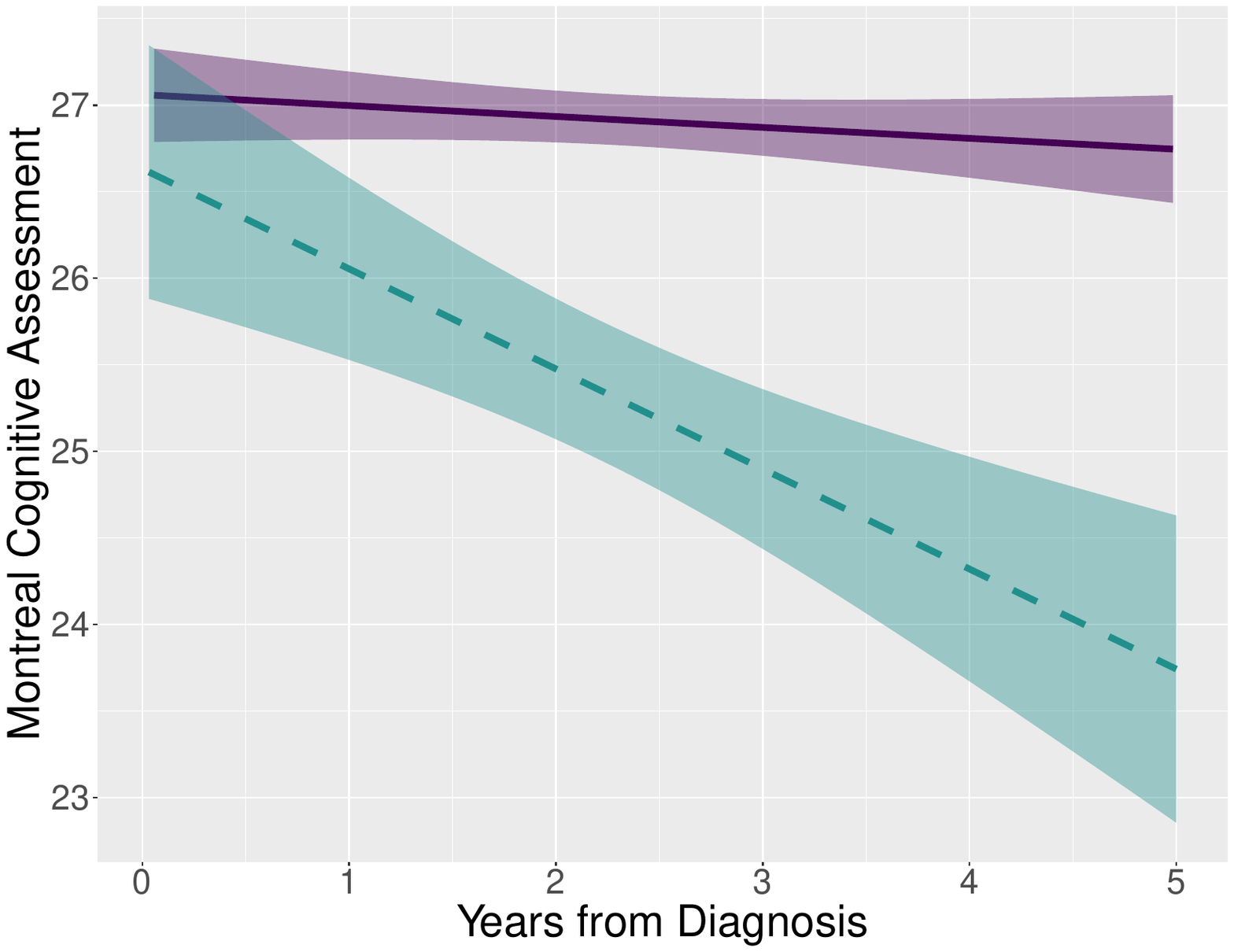}\\
    \caption{Two clusters demonstrated by mean trajectory and 95\% CI of clustering variables: lowest putamen SBR (top left), mean striatum SBR (top right), SCOPA-AUT (second row left), Schwab \& England (second row right), STAI (third row left), GDS (third row right), SFT (bottom), MoCA (bottom right).}
    \label{fig:spagnonclus}
\end{figure}

In order to demonstrate cluster differences in progression beyond those of the clustering variables, we present mean trajectory plots for several non-clustering variables (described below) in Figure \ref{fig:spagnonclus}. The pattern of more rapid worsening is demonstrated by the non-overlapping trajectories which often differ in slope. We quantified the longitudinal differences of the non-clustering variables in the following way. Linear mixed effects models with random intercepts and a linear fixed-effects temporal mean structure were constructed for each longitudinal outcome of interest for the first 5 years since diagnosis per subject. (For imaging variables, this is the first 4.5 years following diagnosis due to limited data beyond this point.) In Table \ref{tab:longvar}, likelihood ratio tests for the interaction between cluster membership and time since diagnosis (i.e. differences in mean slope between clusters) were used to represent differences in progression. If a difference in the slopes was not strongly supported by the evidence, a likelihood ratio test for the additive effect of cluster membership (i.e. one model including a cluster membership main effect and one model without any cluster membership effect) was performed to test any cluster effect.\newline 

Lowest putamen striatal binding ratio (SBR) and mean striatum SBR are common measures of single-photon emission computed tomography (SPECT) imaging of dopamine transporter proteins (DaT) in the brain using the radioligand 123I-ioflupane \citep{tinaz2018semiquantitative}. DaT imaging is used as a diagnostic tool and may be useful monitoring disease progression because loss of nigrostriatal terminals leads to idiopathic PD and parkinsonism and decreases in lowest putamen SBR and mean striatum SBR are associated with worsening clinical features \citep{tinaz2018semiquantitative}. Here both lowest putamen SBR and mean striatum SBR trajectories are lower for the dashed green cluster. The evidence against no cluster-specific slope was weak, but the evidence was stronger against no overall cluster effect. Scales for Outcomes in PD - Autonomic Dysfunction (SCOPA-AUT) measures autonomic symptoms in patients with PD (e.g. symptoms in the gastrointestinal, urinary, sexual, cardiovascular, and respiratory systems; trouble with thermoregulation, and pupillomotor functions), which are common in PD, related to disease duration and severity as well as medication use \citep{visser2004assessment}. Increased SCOPA-AUT is indicative of worse autonomic function. The mean SCOPA-AUT scores are higher and more rapidly increasing with time in the dashed green cluster. Schwab and England activities of daily living (ADL) scale measures independence in patients with PD and other motor disabilities \citep{schwab1969projection}. The scale ranges from 100\%, which is complete independence, to 0\%, which is comatose \citep{schwab1969projection}. The trajectory of Schwab and England ADL scale is worse for the dashed green cluster, dropping below 80\% at which point, it takes an individual twice as long to complete their chores. State trait anxiety inventory (STAI) measures anxiety, and higher scores represent greater anxiety \citep{spielberger2010state}. The STAI for both groups starts at nearly the same value, but improves over time for the solid purple cluster (a diagnosis of PD is very stressful but if progression is slow, perhaps some anxiety dissipates) and worsens for the dashed green cluster which is experiencing worsening symptoms at a greater rate. The geriatric depression scale (GDS) is measure of depression in the elderly, designed to detect depression even in those with other ailments and dementia \citep{sheikh1986geriatric}.  Higher scores indicate more depression with a typical cutoff of at least 5. Mean GDS scores remain stable for the solid purple cluster and increase over time for the dashed green cluster. The semantic fluency test (SFT) is a measure of executive function in which subjects provide as many words as possible from a selected category in a given time \citep{lopes2009semantic}. Higher score represent more words and thus better performance. Mean SFT decreases from time of diagnosis in the dashed green cluster but remains stable in the solid purple cluster. The Montreal cognitive assessment (MoCA) is a brief test of cognitive impairment and dementia, with high scores representing greater impairment \citep{aarsland2021parkinson}. A cutoff score less than 26 is often used as a marker of likely mild cognitive impairment \citep{aarsland2021parkinson}. The dashed green cluster has a steeper decline in MoCA scores with time, which is very strongly supported by the evidence. Considering these variables collectively, the dashed green cluster on average experiences worse trajectories across measures of PD progression. 


\begin{table}[H]
    \centering
    \begin{tabular}{|c|c|c|}
    \hline
        Outcome & $\chi^2 (df = 1)$ & p-value\\
        \hline
        Lowest Putamen SBR & 0.6148 & 0.433\\
        Lowest Putamen SBR\tablefootnote{Due to no strong evidence in support of a cluster by time interaction for lowest putamen SBR, the test of overall cluster effect in main effects model is also presented.} & 7.108 & 0.0077\\
        Mean Striatum SBR & 2.260 & 0.133\\
        Mean Striatum SBR\tablefootnote{Due to no strong evidence in support of a cluster by time interaction for mean striatum SBR, the test of overall cluster effect in main effects model is also presented.} & 15.55 & $<$0.0001\\
        SCOPA-AUT & 17.92 & $<$0.0001\\
        Schwab and England ADL & 26.51 & $<$0.0001\\
        STAI & 11.26 & 0.0008\\
        GDS & 18.20 & $<$0.0001\\
        SFT & 6.669 & 0.009\\
        MoCA & 52.77 & $<$0.0001\\
        \hline
    \end{tabular}
    \caption{Likelihood ratio tests of cluster membership by time interaction in linear mixed effects models of non-clustering ariables. 
    }
    \label{tab:longvar}
\end{table}

The identified clusters did not differ significantly in age; the solid purple and dashed green clusters had mean ages of 60.5 and 61.1 respectively ($p = 0.671$). Compared with the clusters defined by BPE with yearly segments, a total of 8 (2.3\%) observations are clustered differently from this model (BPE with semiannual segments), and parameterizing the survival as Weibull, 12 (3.4\%) observations are clustered differently. 
As an additional sensitivity check, we considered longitudinal models with quadratic and linear time effects. Compared to the employed model (linear time), the model including a term for quadratic time had a total of 9 (2.5\%) observations clustered differently. We also clustered without each of the clustering variables to evaluate cluster stability. There was a great deal of variation in cluster membership for more than two clusters. For two clusters, the following counts (percentages) of observations were clustered differently for each clustering variable excluded: SDMT - 44 (12.4\%), ESS - 17 (4.8\%), LNS - 14 (3.9\%), MDS-UPDRS 1 - 77 (21.7\%), MDS-UPDRS 2 - 26 (7.3\%), MDS-UPDRS 3 - 17 (4.8\%). Despite the larger impacts of excluding MDS-UPDRS 1 or SDMT, clustering using only these two variables results in two clusters with 73 (20.6\%) observations sorted differently than those that we have presented, indicating that the other variables are also important, although weaker, contributors.
\newline

As an important application of this work is clinical trial enrichment, it was practical to determine how well one may predict cluster assignments using enrollment data of variables. The ability to identify the likely subgroup of PD subjects at study enrollment (or during a visit with a physician) would also allow patients, families, and researchers to anticipate an individual's disease progression earlier. To begin this process, a random forest of several variables measured at enrollment and not used for clustering was used to predict cluster membership as described above. Random forest analysis was performed in R with the Machine Shop package \citep{smith}. With 10-fold cross validation, the random forest built without clustering variables had an out-of-bag ROC AUC of 0.747. With the clustering variables at enrollment included, the ROC AUC increased to 0.843. More details about this analysis including ROC curves and variable importance plots are included in the Supplementary Materials.

\section{Conclusion}
\label{sec:conclusion}
In this paper, we developed a cluster analysis method to uncover informative subgroups by differing temporal trends. Unlike existing methods, our approach can form clusters using multivariate longitudinal and time-to-event data, addresses correlations between longitudinal clustering variables, is unambiguous in what variables direct cluster membership, and returns a hierarchy of clusters. \newline

Our motivating application from PPMI identifies clusters with correlated longitudinal assessments and a time-to-event outcome in order to separate PD subjects in subgroups by disease progression. The two clusters identified represent differences in global progression (i.e. decline in all aspects of disease status tested) rather than subtypes defined by a single or a few facets of progression (e.g. motor decline vs. cognitive decline). With the exception of DaT measures, which had cluster-specific intercepts but not cluster-specific slopes, all longitudinal measures differed in progression rate by cluster, with the dashed green cluster progressing more rapidly than the solid purple cluster on average. From the identification of these subgroups, differences in mechanism of disease or biomarkers may be further explored. Further, the identified subgroups can be reasonably predicted at enrollment, making subgroup identification shortly after diagnosis possible. This could be useful for predicting an individual's likely disease trajectory to enrich clinical trials.  \newline

As with all model-based methods, performance is expected to deteriorate to some degree when the model selected does not adequately fit the data. While the presented methods are flexible, allowing for posterior distribution to be specified, our methods are parametric and may perform poorly if the model is misspecified. After clustering, it is recommended to visualize the distributions of variables within clusters to confirm that the distributions specified seem reasonable.  \newline

Within the context of the LMM framework, informative missingness may lead to bias and subsequently misclassifcation. Our model features a conditional independence assumption between data types which may not be satisfied if correlations between data of different types (i.e. a longitudinal variable and a survival variable) are inadequately explained by cluster assignment.  When variables for clustering are dependent on one another within clusters, this assumption is not met and the presented methods may underperform alternative methods which leverage information shared across variables. \newline

Counts of observations per cluster are important to ensure identifiability of the parameters in the LMM. LMM may have multiple random effects and multiple fixed effects. While our framework is flexible, allowing for more complex models (e.g. splines, polynomials, and even covariates), estimating large numbers of parameters requires more data (individuals and/or observations per cluster), placing further restrictions on minimum numbers of observations in each cluster. \newline 

For clustering applications featuring BPE with covariates, if no events occur in two adjacent intervals, then the baseline hazards for those intervals are not identifiable. Additionally, cluster analysis using BPE can be sensitive to changepoints, which can lead to the time-to-event outcome dominating the cluster determination (i.e. our clusters were robust to changepoints of fixed intervals like every 12 or 6 months, but they were sensitive to data driven intervals like time on study quantiles). That is, much care must be used when selecting interval changepoints for BPE models, especially with covariates and/or few events, and sensitivy analyses should be performed. Because of the limitations of the models driving clustering (i.e. number of observations needed to make models identifiable), these methods will fail to identify small clusters, especially if complex models are used. \newline

As with any application involving models, a practitioner must balance model fit with parsimony. The LMM framework relies on identical model matrices for every variable per person. This means that missing data are equivalent for all longitudinal variables at each time and that all variables are subject to the same polynomials, etc. In cases where not all variables are measured at each timepoint or where some variables are likely quadratic in time and others are linear, another option is to assume conditional independence of these longitudinal variables. In this way, model matrices can differ for each variable per person.  \newline

\section{Acknowledgements}
\noindent
Authors would like to thank the Clinical Trials Statistical and Data Management Center at the University of Iowa for their essential input about PPMI data and analyses. \newline\\
\noindent
Data used in the preparation of this article were obtained from the
Parkinson’s Progression Markers Initiative (PPMI) database (www.ppmiinfo.org/data). For up-to-date information on the study, visit www.ppmiinfo.org. PPMI – a public-private partnership – is funded by the Michael J. Fox
Foundation for Parkinson’s Research and funding partners, including PPMI funding partners found at www.ppmiinfo.org/fundingpartners.\newline \\
\noindent
This work was supported in part by Iowa MSTP Training Grant (NIH T32GM007337) and the University of Iowa Ballard and Seashore fellowship.

\bibliographystyle{chicago}
\bibliography{ref}

\end{document}


\begin{flushleft}

\section*{Supplemental Material}

\subsection*{BPE to Poisson Regression}
We will assume that this assumes changepoints are chosen so that no individual's event or censoring time, $t_i$, is a changepoint. Let $N_{ij}$ be an indicator of failure for subject $i$ occurring in the $j^{th}$ interval, $T_{ij}$ be the subject's risk time spent within the interval, and $\delta_{ij}$ be a indicator of the subject's event or censoring occurring in that interval:
\begin{align*}
    N_{ij} &= \mathbf{1}_{[t_i \epsilon (a_{j-1},a_j) \text{ and } d_i = 1]}\\
    T_{ij} &= \begin{cases}
    min(t_i, a_j) - a_{j-1} & t_i  > a_{j-1}\\
     0 & t_i  < a_{j-1}
    \end{cases}\\
    \delta_{ij} &= \mathbf{1}_{[t_i \epsilon (a_{j-1},a_j)]}.
\end{align*}
The likelihood of the BPE model is
    \begin{align*}
         L(\lambda, \beta) &= \prod_i  [f(t_i)]^{d_i}[S(t_i)]^{1-d_i}\\
             &= \prod_i  \bigg[\frac{f(t_i)}{S(t_i)}\bigg]^{d_i}S(t_i)
             = \prod_i [\lambda(t_i)]^{d_i}S(t_i). 
   \end{align*}
Allowing for the flexibility of having covariates $x_i$ and their coefficients $\beta$, 
   \begin{align*}
        \lambda(t_i) &= \lambda_0(t)e^{x_i \beta} = \lambda_0(t)e^{\eta_i}\\
        S(t_i) &= e^{-\Lambda(t_i)} = exp\{-e^{\eta_i}\int_0^{t_i}\lambda_0(u)du\}\\
        L(\lambda, \beta) &= \prod_i  [\lambda_0(t)e^{\eta_i}]^{d_i}exp\{-e^{\eta_i}\int_0^{t_i}\lambda_0(u)du\}
    \end{align*}
Working with the likelihood for an individual,
\begin{align*}
        L_i(\lambda, \beta) &=  [\lambda_0(t)e^{\eta_i}]^{d_i}exp\{-e^{\eta_i}\int_0^{t_i}\lambda_0(u)du\}\\
        &= \prod_j [\lambda_j e^{\eta_i}]^{N_{ij}}exp\{-\delta_{ij}[\lambda_j(t_i - a_{j-1}) + \sum_{g=1}^{j-1} \lambda_g (a_g - a_{g-1})] e^{\eta_i}\}.
\end{align*}
Reasonably, where $\delta_{ij}= 0$ (no event/censoring occurs in interval j), there is no contribution to the likelihood, $L_i(\lambda) = 1$. Alternatiavely, if $\delta_{ij}= 1$ and $d_i = 1$ (there is an event in interval $j$),the contribution is hazard $\times$ survival. If $\delta_{ij}= 1$ and $d_i = 0$ (there is an censoring in interval $j$), the contribution is survival. Putting that together, for a subject who was either censored or experienced an event in interval j, $\delta_{ij} = 1$,
\begin{align*}
        L_i(\lambda, \beta) &=[\lambda_j e^{\eta_i}]^{N_{ij}}exp\{-[\lambda_j(t_i - a_{j-1}) + \sum_{g=1}^{j-1} \lambda_g (a_g - a_{g-1})]
        e^{\eta_i}\}\\
        &=[\lambda_j e^{\eta_i}]^{N_{ij}}exp\{- \sum_{g=1}^{j} \lambda_g H_{ig}
        e^{\eta_i}\}\\
        &=[\lambda_j e^{\eta_i}]^{N_{ij}}[\prod_{g=1}^{j}exp\{- \lambda_g H_{ig}
        e^{\eta_i}\}][\prod_{g=1}^{j-1}(\lambda_g H_{ig}
        e^{\eta_i}\})^0]\\ 
        &\propto [\lambda_j T_{ij} e^{\eta_i}]^{N_{ij}}[\prod_{g=1}^{j-1}(\lambda_g H_{ig}
        e^{\eta_i}\})^0] [\prod_{g=1}^{j}exp\{- \lambda_g H_{ig}
        e^{\eta_i}\}]\\
        &= \prod_{g=1}^{j}[\lambda_g H_{ig}
        e^{\eta_i}\}]^{N_{ig}} exp\{- \lambda_g H_{ig}
        e^{\eta_i}\}\\
        &= \prod_{g=1}^{j}\frac{[\lambda_g H_{ig}
        e^{\eta_i}\}]^{N_{ig}} exp\{- \lambda_g H_{ig}
        e^{\eta_i}\}}{N_{ig}!}.
\end{align*}
Recognizing the kernel of a Poisson distribution, we have
\begin{align*}
        N_{ig} &\sim Pois(\mu_{ig} = \lambda_g H_{ig}
        e^{\eta_i})\\
        \log(\mu_{ig}) &= \log(H_{ig}) + \log(\lambda_{g}) + \eta_i
    \end{align*}
This is Poisson regression with offset $\log(H_{ig})$ and interval specific intercept $\log(\lambda_{ig})$. The outcome $N_{ig}$, indicator of event occurring for individual $i$ in interval $g$ is clearly not independent within an individual (all zero except the last one, which is either zero or one). The form of the likelihood is a product of interval-specific Poisson contributions.\newline 

Because we are not using covariates, the MAP for $\lambda$ is in closed form by using conjugate priors $\lambda \sim \Gamma(a,b)$,
    \begin{align}
    \pi(\lambda_g|y) &= \bigg[\prod_i (\lambda_g H_{ig} )^{N_{ig}}e^{\lambda_g H_{ig}}\bigg] \frac{b^a}{\Gamma(a)}\lambda_g^{a-1}e^{-b\lambda_g} \\
    \lambda_g|y &\sim \Gamma(a + \sum_i N_{ig}, b + \sum_i H_{ig})\\
    \hat{\lambda_g}_{MAP} &= \frac{a - 1 + \sum_i N_{ig}}{b + \sum_i H_{ig}}
    \end{align}
The form of the MAP is intuitive; the numerator involves the count of individuals who experienced an event in interval g and the denominator involves the sum of time at risk in the interval across individuals.

\subsection*{Random Forest Methods and Results}
To formally quantify the extent to which clusters based on temporal trends can be predicted using data collected at enrollment, we use random forest algorithms. A random forest model was selected because it provides the flexibility of classification trees (numeric and categorical data with minimal preparation, considers interactions) with the strength of ensemble learning (improved accuracy), and relatively simple tuning. Included variables are displayed below in the variable importance plots. Within 10-fold cross validation, model parameters were tuned over the number of random features considered at each split using out-of-bag error rate. The ROC AUC was 0.747 for predicting cluster membership using enrollment data excluding variables used to cluster and 0.843 including clustering variables. This indicates that it may be possible to predict which subjects/patients are likely to have more rapid progression using data collected within a couple of years of diagnosis. The most important variables for determining cluster membership in the random forest without clustering variables were the REM behavior disorder (RBD) questionnaire, SCOPA-AUT score, tau amyloid beta  ratio, semantic fluency test score, alpha synuclein, and serum neurofilament light chain (NfL). RBD is a common symptom of PD and its prodrome. SCOPA-AUT measures of autonomic function and is further discussed in the main text. CSF abeta forms neurotoxic plaques and CSF tau is an axonal microtubule protein. Lower CSF abeta have been shown to be associated Alzheimer disease and  memory impairment \citep{alves2010csf}. Studies demonstrate that higher levels of CSF tau, which is correlated with alpha synuclein , are associated with motor progression \citep{hall2015csf}. Alpha synuclein aggregates are a hallmark of PD and other synucleinopathies. Serum NfL, a biomarker of axonal damage, and is similarly correlations with alpha synuclein \citep{hall2015csf}. For the random forest including clustering variables, MDS-UPDRS 2 was by far the most important variable with MDS-UPDRS 1, SDMT, ESS, and MDS-UPDRS 3 also having large importances. Similar results, both in performance and variable importance, were obtained using gradient boosting.

\begin{figure}[H]
    \centering
    \includegraphics[scale = .5]{longitudinal/PPMI Figures/roc1.pdf}\\
    \includegraphics[scale = .5]{longitudinal/PPMI Figures/roc2.pdf}\\
    \caption{Random Forest ROC Curve for Predicting Cluster using Baseline Variables, excluding clustering variables (top) and including clustering variables (bottom). }
    \label{fig:roc}
\end{figure}

\begin{figure}[H]
    \centering
    \includegraphics[scale = .45]{longitudinal/PPMI Figures/vi1.pdf}\\
    \includegraphics[scale = .45]{longitudinal/PPMI Figures/vi2.pdf}
    \caption{Random Forest Variable Importance for Predicting Cluster using Baseline Variables, excluding clustering variables (top) and including clustering variables (bottom).}
    \label{fig:vi}
\end{figure}

\bibliographystyle{chicago}
\bibliography{ref}

    \end{flushleft}